\documentclass[fleqn,usenatbib]{mnras}
\usepackage[T1]{fontenc}
\usepackage{xcolor}
\usepackage{amsmath}
\usepackage{graphicx}
\newcommand{\eb}{\begin{equation}}
\newcommand{\ee}{\end{equation}}

\definecolor{rkka}{RGB}{219,66,32}

\title[Mapping 3D Rotation States of Triaxial Celestial Bodies]{Chaos over Order: Mapping 3D Rotation of Triaxial Asteroids and Minor Planets}

\author[V. V. Makarov et al.]{
Valeri V. Makarov,$^{1}$\thanks{E-mail: valeri.makarov@gmail.com}
Alexey Goldin,$^{2}$\thanks{E-mail: alexey.goldin@gmail.com}
Alexei V. Tkachenko$^{3}$
Dimitri Veras$^{4,5,6}$
Beno{\^i}t Noyelles$^{7}$
\\
$^{1}$United States Naval Observatory, 3450 Massachusetts Ave. NW, Washington, DC 20392-5420, USA\\
$^{2}$Teza Technology, 150 N Michigan Ave, Chicago IL 60601, USA\\
$^{3}$Center for Functional Nanomaterials, Brookhaven National Laboratory, Upton NY 11973,   USA\\
$^{4}$Centre for Exoplanets and Habitability, University of Warwick, Coventry CV4 7AL, UK\\
$^{5}$Centre for Space Domain Awareness, University of Warwick, Coventry CV4 7AL, UK\\
$^{6}$Department of Physics, University of Warwick, Coventry CV4 7AL, UK\\
$^{7}$Institut UTINAM, UMR 6213 / CNRS, Univ. Bourgogne Franche-Comt\'e, OSU THETA, BP 1615, 25010 Besan\c{c}on Cedex, France\\
}
\pubyear{2022}
\date{Accepted XXX. Received YYY; in original form ZZZ}
\begin{document}
\label{firstpage}
\pagerange{\pageref{firstpage}--\pageref{lastpage}}
\maketitle

\begin{abstract}
Celestial bodies approximated with rigid
triaxial ellipsoids in a two-body system can rotate chaotically due to the time-varying gravitational torque from the central mass. At small orbital eccentricity values, rotation is short-term
orderly and predictable within the commensurate spin-orbit
resonances, while at eccentricity approaching unity, chaos completely takes over. Here, we present the full 3D rotational equations of motion around all three principle axes for triaxial minor planets and two independent methods of numerical solution based on Euler rotations and quaternion algebra. The domains of chaotic rotation are numerically investigated over the entire range of eccentricity with a combination of trial integrations of Euler's equations of motion and the GALI($k$) method. We quantify the dependence of the order--chaos boundaries on shape by changing a prolateness parameter, and find that 
the main 1:1 spin-orbit resonance disappears for specific moderately prolate shapes already at eccentricities as low as 0.3. The island of short-term stability around the main 1:1 resonance shrinks with increasing eccentricity at a fixed low degree of prolateness and completely vanishes at approximately 0.8. This island is also encroached by chaos on longer time scales indicating longer Lyapunov exponents. Trajectories in the close vicinity of the 3:2 spin-orbit resonance become chaotic at smaller eccentricities, but separated enclaves of orderly rotation emerge at eccentricities as high as 0.8. Initial perturbations of rotational velocity in latitude away from the exact equilibrium result in a spectrum of free libration, nutation, and polar wander, which is not well matched by the linearized analysis omitting the inertial terms.
\end{abstract}
\begin{keywords}
chaos -- methods: numerical -- minor planets, asteroids: general -- celestial mechanics -- planet–star interactions
\end{keywords}

\section{Introduction}
Current interest in and scrutiny of Solar system minor bodies --- moons, asteroids and comets --- is unprecedented. For example, the {\it Cassini} mission provided us with detailed shape models of small Saturnian satellites \citep{thomas2010,spilker2019}, and the recently launched {\it DART} mission will generate the most physical intimate portrait yet of another type of moon, one orbiting an asteroid \citep{agretal2021,rivetal2021}. Fundamental, but still not fully understood, properties of these minor planets are their spin orientation and evolution.

Spin evolution also plays a critical role in non-resonant asteroids. In fact, the size distribution of asteroids in the inner Solar system and the debris profile of this region are largely driven by spin-induced breakup through the radiative YORP effect \citep{botetal2006,voketal2015,huetal2021}. As asteroids spin up or down due to the Sun's radiation, their spin states and orientations might change in unexplored ways due to their shape, affecting which types of asteroids break up and when. Further, the importance of rotational breakup characteristics is not limited to the Solar system, with consequences for shaping planetary debris distributions in white dwarf planetary systems \citep{makver2019,2020MNRAS.492.5291V}.

One context in which minor body spin is relatively well understood are
planetary satellites in the Solar system which are locked in spin-orbit resonances. These satellites rotate in an orderly fashion with spin rates commensurate with the orbital frequencies. The commonly observed resonance is 1:1, i.e., synchronous rotation, with the Earth-Moon pair being the best-known example.

The main reason for this prevalence of order is the range of orbital eccentricity $e$, which tends to be low. Chaotic rotation may arise from overlap of different spin-orbit resonances \citep{chi79}. As shown by \citep{wis84}, the half-width of the chaotic separatrix between the synchronous and the 3:2 spin-orbit resonances is
\begin{equation}
    w_1 = \frac{14\pi e}{\omega_0^3}\exp\left(-\frac{\pi}{2\omega_0}\right),
\end{equation}
where $\omega_0 = 3n(B-A)/C$ is the asymptotic frequency of the free libration of the resonant argument in a one-dimensional (1D) approximation, $n$ is the orbital mean motion, and $A < B < C$ are the moments of inertia of the ellipsoidal satellite. Conditions of rotational chaos in the full 3D setup have been much less investigated.

A commensurate spin-orbit state is a stable resonance and an attractor in
the parameter space due to a restoring force, which counteracts any perturbation directed away from the point of equilibrium. If the orbit is significantly elongated, most of the action takes place around the pericenter because of the explicit proportionality of the spin acceleration to $(a/r)^3$, where $a$ is the semimajor axis and $r$ is the instantaneous distance between the bodies. \citet{wis87} discussed interesting cases where this pericenter interaction becomes too strong to keep the trajectory within the resonance zone. The strength of the pericenter interaction depends on the eccentricity and, to a lesser degree, on the degree of elongation. Hyperion, with its significant eccentricity (0.1) pumped by Titan, was historically the first example of departure from this scene ruled by order \citep{wis84}. Although its 1:1 resonance is present, the surrounding chaotic zone makes it attitude-unstable. Other resonances, such as 3:2, are completely removed by chaos.  Based on theoretical analysis of chaos conditions and numerical experiments, \citet{kou05} added Prometheus and Pandora to the list of objects with disorderly rotation. Finally, analysis of a more representative sample of satellites and minor bodies revealed that chaos is a norm rather than a rarity in solar system rotational dynamics \citep{mel10}.

Most of the previous analyses focused on the planar case of the principal axis of rotation being always orthogonal to the orbital plane. The equation of motion becomes one-dimensional (1D) with the gravitational torque aligned with the axis of the greatest moment of inertia. Ignoring the contribution of the satellite's mass, which is much smaller than the planet's mass, the well known equation of motion in this toy model in the inertial frame \citep[e.g.,][]{gol} can be written as
\eb
\Ddot{\theta}=-\frac{3}{2}n^2{\left(\frac{a}{r}\right)}^3\frac{B-A}{C}\sin 2(\theta-f),
\label 
{1D.eq}
\ee
where $n$ is the orbital mean motion (i.e., frequency), $a$ is the semimajor axis of the orbit, $r$ is the instantaneous separation between the bodies, $f$ is the true anomaly, and $\theta$ is the orientation angle of the longest satellite's axis with respect to the line of apsides. The periodic driving force in the right-hand part can be decomposed into an infinite series of harmonics of the mean anomaly with coefficients that are certain combinations of the Hansen's coefficients, i.e., functions of eccentricity \citep{1964RvGSP...2..661K}. The emerging equation is that of a driven non-harmonic pendulum without damping (if we ignore tidal friction). In the approximation of sufficiently small libration amplitudes in the close vicinity of a spin-orbit resonance, the solution for a harmonic driven oscillator is often used \citep{fro}. Numerical integrations of the 1D equation of motion show that this approximate analytical solution is fairly accurate for a small eccentricity and triaxiality parameter $(B-A)/C$, with the exception of the initial conditions required to remove free libration, which are sensitive to the nonlinearity of the driving torque with respect to the libration angle. 

This comfortable and intellectually rewarding situation breaks up when one advances to the full 3D case where the rotating body is allowed to move around all three principle axes of inertia. Even a restricted model with only one degree of freedom in rotation but a finite tilt of the rotation vector to the orbital plane is prone to dynamical chaos \citep{kwi}.
The conditions for chaotic rotation and the characteristic Lyapunov times
are known to depend on the degree of elongation of triaxial bodies. 
Even in the generally more stable 1D setup, an additional regularization
agent is required to make Hyperion's rotation stable and predictable
\citep{tar}.

The non-inertial part of the equations of motion is defined by a time-variable, nonlinear gravitational torque from the primary body, which is also unequally distributed between the three dimensions. These nonlinear dependencies are at the origin of the emerging chaotic behaviour. The base model, the range of eccentricity, and the shape parameter  used in this paper are described in Sect. \ref{ecc.sec}. We use the GALI($k$) numerical method to probe for chaos, which is briefly introduced in Sect. \ref{gali.sec}. This method requires fast and accurate algorithms of integration of the three Euler's equations of motion in the inertial body frame, which are coupled with the attitude reconstruction problem in the fixed inertial frame attached to the orbit (world frame), as discussed in Sect. \ref{int.sec}. We solve the attitude problem using two different techniques, a commonly used 3-2-1 sequence of Euler rotations and a more algebraic quaternion implementation, which are detailed in Appendices A and B, respectively. Both techniques produce the same results for stable trajectories with identical initial conditions, but the quaternion implementation is our preferred numerical method. Using a model body with parameters similar to Enceladus, the Saturn's moon, we perform test integrations with varying initial conditions in the vicinity of the 1:1 spin-orbit resonance and compute the frequency power spectra (periodograms) of the resulting 3D free libration in the body frame, nutation in the world reference frame, and the polar wander in the world frame (Sect. \ref{fre.sec}). The main results are presented in Sect. \ref{cha.sec}, where we discuss the numerically constructed GALI(2) maps of chaos versus order in the cross-section of eccentricity and shape parameter, as well as the shape and size of the islands of short-term stability centred on the 1:1 and 3:2 resonances, and their erosion and disappearance with increasing eccentricity. The main results are summarised in Sect. \ref{sum.sec}, where we also discuss the outlook for further studies and remaining open issues.

\section{Eccentricity, shape, and gravitational torque}\label{ecc.sec}

In this paper, we consider a class of celestial bodies whose inertia tensor can be well approximated with a diagonal tensor
$\boldsymbol I={\rm diag}\{A,B,C\}$ in a Cartesian coordinate system $\{\boldsymbol Y_1,
\boldsymbol Y_2,\boldsymbol Y_3\}$. This class includes all objects with symmetric inertia tensors, because an orthogonal (and unitary) transformation of coordinates can be found to diagonalize such tensors. The values $A\le B\le C$ are the moments of inertia, and
the basis vectors $\{\boldsymbol Y_1,
\boldsymbol Y_2,\boldsymbol Y_3\}$ define the principal axes of inertia. We broadly call such objects triaxial bodies. A simple example is a solid ellipsoid with semimajor dimensions $a$, $b$, and $c$. The shape of an ellipsoid is fully described by two
parameters $s_1$ and $s_2$ with the geometric equation \citep
{1980A&A....82..289B}
\eb 
x_1^2/s_1^2+x_2^2/s_2^2+x_3^2=c^2.
\ee 
with $s_1\ge s_2\ge 1$. Special cases are a prolate figure at
$s_2=1$ and an oblate figure at $s_2=s_1$. The correspondence to the aspect
ratios used by \citet 
{2020MNRAS.492.5291V} is $s_2=\beta$, $s_1=\gamma$. The equations of 3D
rotation (Appendix A) only involve the ratios of the principal moments of
inertia. Therefore, it is sufficient to compute the scaled analogs of these
moments. For an ellipsoid of uniform density, the scaled inertia moments
can be written as $A=s_2^2+1$, $B=s_1^2+1$, and $C=s_1^2+s_2^2$. One of the
objectives for this study is to estimate how chaotic rotation can emerge for
celestial bodies of different degrees of asphericity. To reduce the number
of degrees of freedom to 1, we introduce a figure prolateness parameter $L_f$, which scales the relative extension along the $a$ semiaxis by greater amounts than along the $b$ semiaxis keeping the $c$ dimension unchanged. The scaled shape parameters are
\begin{eqnarray}
s_1 &=& (a/c-1)\,L_f+1 \nonumber \\
s_2 &=& (b/c-1)\,L_f+1.
\end{eqnarray}

Most of our computations presented in this paper are performed for a model
object with variable eccentricity and prolateness $L_f$ but fixed initial
dimensions $a=256.3$, $b=247.3$, $c=244.6$, which, expressed in km, correspond to the size of Saturn's moon Enceladus. The motivation for this choice was our interest in the past rotational history of Enceladus during possible episodes of a higher orbital eccentricity, as well as the fact the Enceladus is close to a perfectly prolate shape. The small $b/c$ ratio is also a conscious choice to focus on almost perfectly prolate shapes, which may be more relevant for asteroids. With these dimensions, the initial inertia coefficients (at $L_f=1$) in the Euler equations of motion (\ref{eu.eq})
are $(B-A)/C=0.03573$, $(C-A)/B=0.04669$, and $(C-B)/A=0.01098$. These small to moderate values are characteristic of the larger moons in the Solar system, while asteroids are expected to often be much more deformed.

The present day eccentricity of Enceladus is $0.0047$, which is remarkably small. On the other hand, main belt asteroids have much greater eccentricities, with one outstanding case, 2006 HY51, investigated in 1D by \citet  {2020ApJ...899..103M} for the statistical parameters of its chaotic rotation. The previous 1D studies have revealed that eccentricity is indeed a critical
parameter separating the domains of stable and chaotic rotation. We develop here a computing algorithm for the full 3D setup that is numerically stable and accurate for the entire range of $e\in [0,0.99]$. This methodology can therefore be used for a broad range of celestial objects in the Solar system and outside. Nereid, the distant moon of Neptune, is an interesting possible application with its high eccentricity ($e=0.7417$) and unknown
shape, whose 3D rotation can be mapped for chaos. Nereid's rotation rate is apparently high, so unless it is chaotically tumbling, its principal axis
$\boldsymbol Y_3$ orientation is likely to drift with respect to the ecliptic with a rate of up to $\sim 0.01$ rad per century. The only celestial body with a certainly known chaotic rotation with a short Lyapunov time is Hyperion \citep  
{2015Ap&SS.357..160T}. The distinction probably comes from its strongly elongated shape, which suggests large coefficients of inertia, $(B-A)/C=0.29$, $(C-A)/B=0.51$, and $(C-B)/A=0.25$, hence, much stronger torques then for Enceladus. Hyperion's eccentricity, on the other hand, is rather moderate at 0.123. We surmise from these examples that manifestly chaotic rotation of triaxial bodies happens at certain combinations of eccentricity and prolateness, both tending to the higher ends of their ranges.

\section{Detection of chaos by GALI}
\label{gali.sec}

Generalized Alignment Index (GALI) \citep 
{2014arXiv1412.7401S} is one of the numerical methods to detect chaotic behaviour of a dynamical system. The advantage of this method over more commonly known techniques such as the Poincar{\'e} surface of section mapping is a numerically economic way of estimation on a metric limited to a single parameter, which allows its application in problems of high dimensionality \citep 
{2020arXiv200700941T}. In our case, this is a decisive advantage because we are investigating a vast parameter space. The basic idea of the method is to generate reasonably small initial perturbations to a specific trajectory (i.e., a particular integrated solution with fixed initial conditions) and check their mutual alignment within a certain characteristic time interval $t_{\rm max}$. The initial perturbation vectors should be orthogonal. For a chaotic parent trajectory, the originally orthogonal differences converge in the direction of a Lyapunov exponent and become nearly aligned. The formal verdict is the parent trajectory is chaotic if the GALI($k$) index falls below a threshold value $\nu_{\rm lim}$ within $t_{\rm max}$. The number of perturbation vectors $k$ cannot be greater than the rank of the parameter space, which is 6 in our case (three angular coordinates and three angular velocities). We used a very low threshold of $10^{-20}$ and, in the massive simulations, $t_{\rm max}=275$--1000 orbits capitalizing on the fast and stable performance of the published Julia code\footnote{\url{https://juliadynamics.github.io/DynamicalSystems.jl/v1.3/chaos/chaos\_detection/}} \citep{dat}.

Technically, the GALI method is similar to the somewhat less quantitatively definite method of sibling trajectories separation based on the exponential divergence of initially very close chaotic trajectories. This class of techniques has been used to investigate the multi-dimensional parameter space of the Solar system \citep 
{2008MNRAS.386..295H, 2007NatPh...3..689H}, detect chaotic orbits in exoplanet systems \citep 
{2014ApJ...780..124M}, and to map the stability areas of generally chaotic motion of high-eccentricity and high-inclination orbits of exoasteroids orbiting white dwarfs \citep 
{2019A&A...629A.126A}. For this paper, we mostly used $k=2$ because it seems to be sufficient to capture the fine detail of the chaos/regular transition zones surrounding the islands of stability.

\section{Integrating equations of motion}
\label{int.sec}

We introduce an inertial, non-rotating with respect to distant stars,
reference frame defined by the stationary orbit of the celestial body
around a central mass, which is considered to be a point mass in this
paper. The coordinate triad $\{\boldsymbol X_1,\boldsymbol X_2,\boldsymbol X_3\}$ is defined so that ${\boldsymbol X}_1$ is directed
toward the pericentre, ${\boldsymbol X}_3$ is aligned with the orbital
angular momentum, and ${\boldsymbol X}_2$ completes the right-handed
coordinate frame. With respect to the ellipsoidal body, the gravitational
attractor moves with a varying velocity within the plane spanned by
${\boldsymbol X}_1$ and ${\boldsymbol X}_2$. We consider another reference
frame firmly attached to the rotating body, defined by the right-handed
triad of basis vectors $\{\boldsymbol Y_1,\boldsymbol Y_2,\boldsymbol Y_3\}$. Here the axes are aligned with the principal axes of inertia
of the ellipsoidal body $A$, $B$, and $C$, respectively, where $A\leq B\leq C$. The well-known Euler's equations of motion \citep{1962fcm..book.....D} refer to this latter system and can be written in the form:
\begin{eqnarray}
\ddot y_1 & = & \frac{C-B}{A}\left[\frac{3GM}{r^3}h_2h_3-\dot y_2\dot y_3
\right] \nonumber\\
\ddot y_2 & = & \frac{A-C}{B}\left[\frac{3GM}{r^3}h_3h_1-\dot y_3\dot y_1
\right] \nonumber\\
\ddot y_3 & = & \frac{B-A}{C}\left[\frac{3GM}{r^3}h_1h_2-\dot y_1\dot y_2
\right] 
\label{eu.eq}
\end{eqnarray}
where $[\dot y_1,\dot y_2,\dot y_3]$ is the instantaneous rotation vector
in the body frame, $[\ddot y_1,\ddot y_2,\ddot y_3]$ is the instantaneous
angular acceleration vector, $G$ is the gravitational constant, $M$
is the mass of the gravitating body, $r$ is the instantaneous distance
between the two interacting bodies, and $h_i, i=1,2,3$, are the direction
cosines of the vector from the center of the triaxial body to the gravitating point mass in the instantaneous body frame. The gravitational part of the
torque can be derived from the general equation involving an inertia matrix
\citep[][p. 190]{2009amss.book.....S},
taking the simplest diagonal form in the body frame.

The free rotation (inertial) part of these equations of motion is derived directly from the two conservation
laws \citep 
{2012omsp.book.....S},  the total kinetic energy 
\eb
2\,E=A\dot y_1^2+B\dot y_2^2+C\dot y_3^2
\ee 
and the angular momentum squared
\eb 
L^2=A^2\dot y_1^2+B^2\dot y_2^2+C^2\dot y_3^2.
\ee
These quantities are not constant in any other frames including the
orbit-fixed inertial frame.
It is important for the following derivation to understand the properties
of the body frame where Eqs. \ref{eu.eq} are valid. It is the inertial
(non-rotating) frame which is instantaneously aligned with the principal
axes of inertia at a given time $t$. The physical process of rotation
can be viewed as a continuous series of inertial body frames forming a
sequence of infinitesimal 3D rotations.

The second-order ODEs in Eq. \ref{eu.eq} can be directly integrated
with a proper set of initial conditions to yield the velocity
vector components $\dot y_i$ as functions of time. The main technical
difficulty is computing the direction cosines $s_i$ as functions of time.
These are easily computed in the inertial frame $\{\boldsymbol X_1,\boldsymbol X_2,\boldsymbol X_3\}$, but we need to know the direction
to the perturber at any time in the body frame. It would seem that
the integrated angles $y_i$ should define the orientation of the body
frame with respect to the orbital frame, but in fact, this is not
the case. These angles are some integrated angular paths  of the body
with respect to the initial orientation, which have no immediate
physical interpretation. We have therefore a classical attitude reconstruction problem to solve: given a set of initial conditions and
2nd order ODEs, determine the mutual orientation of the two frames in
question. 

In this paper, we use two essentially different techniques to solve the attitude problem, viz., a standard 3-2-1 sequence of Euler rotations (Appendix A) and a quaternion representation (Appendix B). Although the latter approach requires one more ODE to be solved, we found it to be faster and more numerically accurate. Variations of the former technique, on the other hand, have been traditionally used in such cornerstone problems of celestial mechanics as description of Lunar rotation and orientation. \citet 
{2011CeMDA.109...85R} describe a rather involved 3-1-3-1 schema of Euler rotations traditionally used for the Lunar attitude, which has explicit numerical singularities. As long as the motion of the Moon is regular and confined to a narrow torus in the parameter space, these degeneracies are of no concern. Our 3-2-1 schema also has a singularity related to the second rotation, which becomes an issue for a tumbling body. Otherwise, the two methods obtain practically identical results for regular trajectories with identical initial conditions. As a sanity check, we integrated our basic Enceladus model (with $n=4.585537$ d$^{-1}$, $\sigma=(B-A)/C=0.03573$, $e=0.0047$) with the initial conditions of ideal synchronisation (at pericentre time $t=0$, all attitude angles equal to zero, and spin rates are $\omega_1(0)=\omega_2(0)=0$, $\omega_3(0)=0.99885\,n$)\footnote{The small offset of the rotation rate from $1\,n$ at pericentre is due to the harmonic pendulum approximation of the amplitude of forced libration in longitude, $-6\,e\, \sigma \,n^2/({\cal N}^2 - n^2)$, where ${\cal N}$ is the natural frequency of free libration in longitude, cf. \citep{fro}.}
and obtained with both methods a purely longitudinal forced libration with the theoretically expected from Eq. \ref{1D.eq} asymptotic amplitude $6\, e\, \sigma/(3\sigma-1)=-0.0647\degr$.

\begin{figure*}
\centering
\includegraphics[width=.32\textwidth]{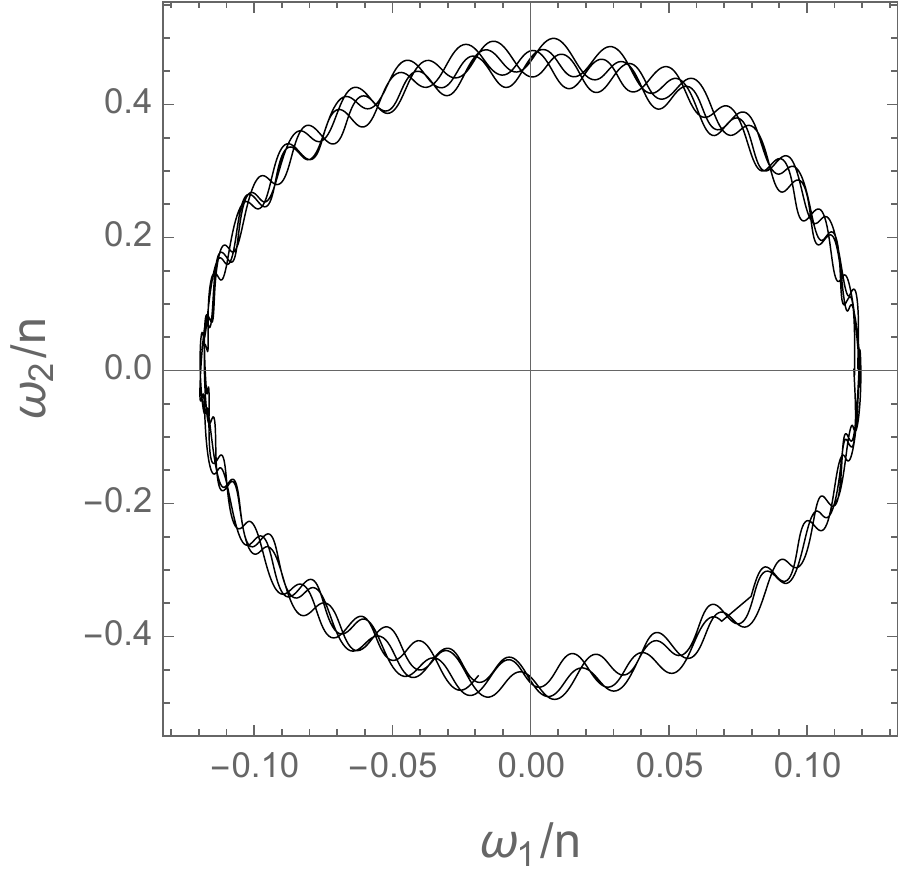}
\includegraphics[width=.32\textwidth]{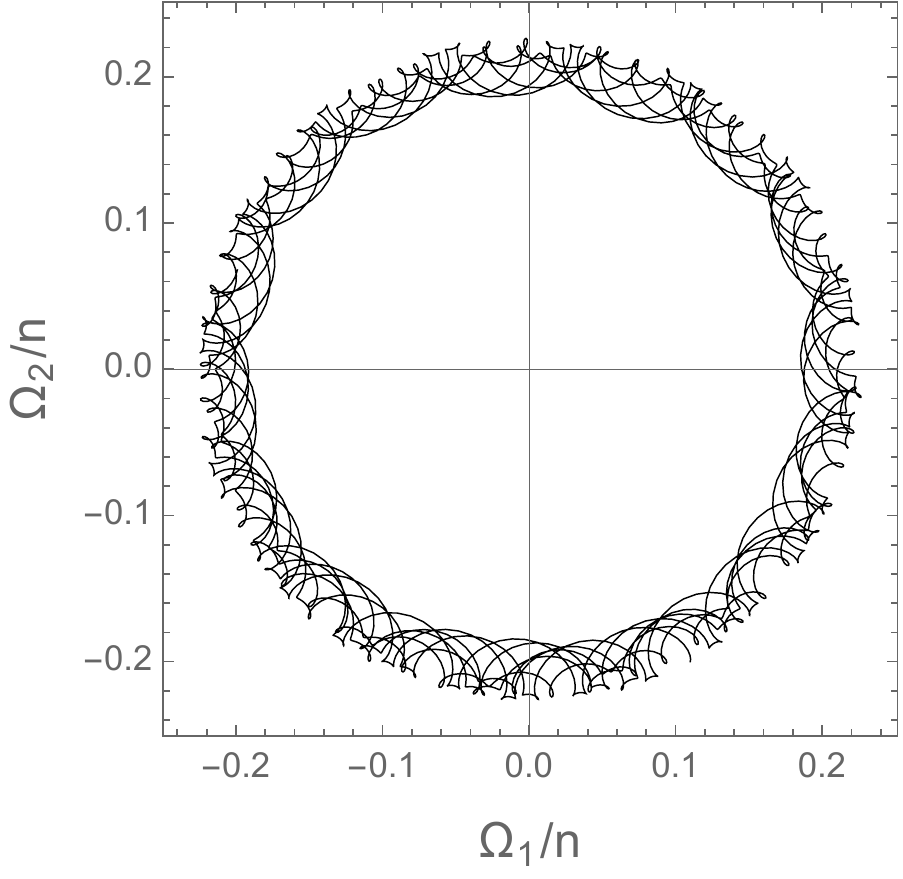}
\includegraphics[width=.32\textwidth]{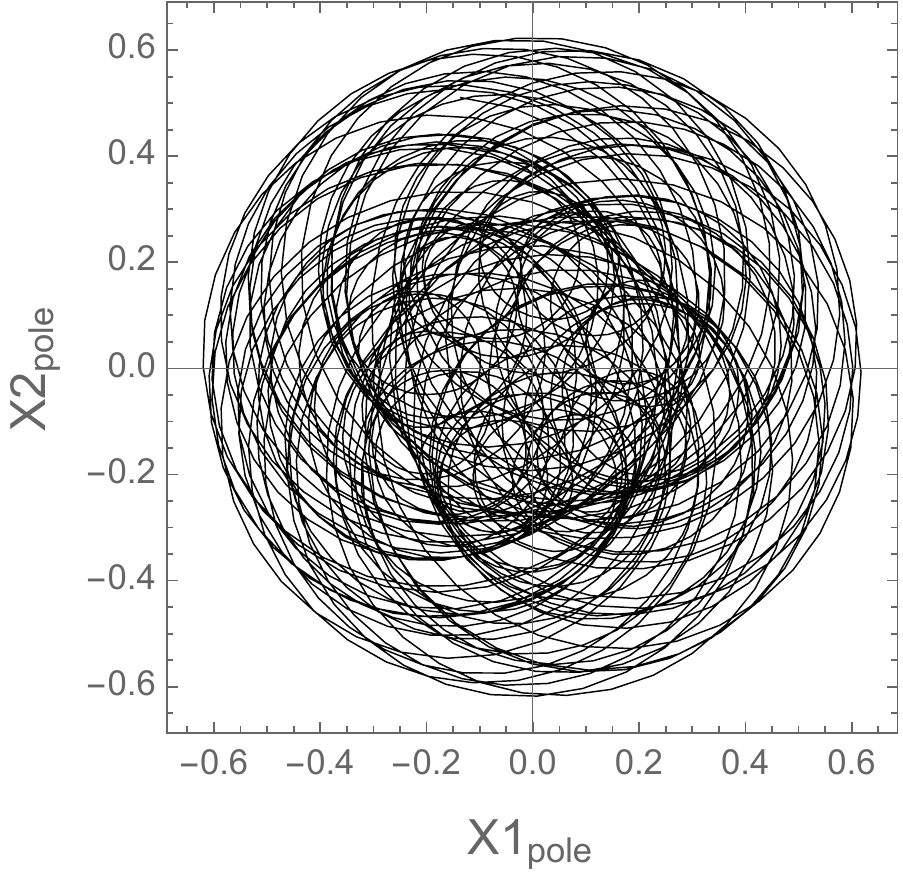}
\caption{Simulated precession, nutation, and pole wander of a test body with
$L_f=1$, $e=0.0047$, $n=4.5855$ d$^{-1}$ (model Enceladus) and initial
conditions described in the text. Left: normalized roll and pitch components of
rotation rate in the body frame; middle: normalized components of rotation
rate projected onto the inertial plane of orbit; right: coordinates of the true
pole $\boldsymbol r=[0,0,1]$ in the plane of orbit. \label{pole.fig}}
\end{figure*}

With these tools in hand, we can accurately and swiftly integrate the motion of our base model Enceladus with any initial conditions. As soon as we depart from the perfect synchronism, significant free librations emerge making the rotational behaviour quite complex. Fig. \ref 
{pole.fig} shows the solution for heavily perturbed initial conditions of the base model and $L_f=1$, which corresponds to $(B-A)/C=0.03573$, $(C-A)/B=0.04669$, and $(C-B)/A=0.01097$.
The initial periastron orientation remains aligned with the inertial orbit frame (zero initial roll, pitch, and yaw angles) and the initial velocities are $\omega_1(0)=0.11\,n$, $\omega_2(0)=0.2\,n$, $\omega_3(0)=1.0 \,n$. The angular velocity vector loosely describes a cone both in the body frame (left plot) and the orbital frame (centre plot), which corresponds to the free libration. This relatively slow quasi-periodic motion is superimposed with a higher-frequency wobble due to the forced libration, or nutation. The vector $[0,0,1]$ in the body frame defining the north pole of the body describes high-amplitude loops in the inertial frame remaining within $\sim 0.6$ radians of the orbit's axis. These are the signs of a regular trajectory around the 1:1 spin-orbit resonance with a large free libration component.

More complex patterns of free libration may emerge for trajectories in the vicinity of higher-order spin-orbit resonances, e.g., the 2:1 resonance. The velocity vector describes wide loops that are not confined to a certain cone in either reference frame. The trajectory cannot be called quasi-periodic in any sense. The longitudinal libration in the body frame shows a few beating oscillations of apparently non-commensurate frequencies, which makes the trajectory, being non-chaotic and regular, to never return to the same point in the 6D parameter space.

\section{Characteristic frequencies of 3D rotation}
\label{fre.sec}
The trajectory projections shown in Fig. \ref{pole.fig} for a case of model Enceladus with strongly perturbed initial velocities in the latitude dimensions show both free and forced libration patterns. Even in the simplest 1D case, both the amplitude and frequency of free libration depend on the initial conditions, or, in general, on the excess of kinetic energy \citep{2015ApJ...810...12M}. The equations of motion (\ref{eu.eq}) are explicitly nonlinear and there is no obvious way of computing the eigenfrequencies without simplifying assumptions. The commonly used simplification even in the most advanced theory of Lunar rotation is to ``linearize" the equations of motion by ignoring the inertial terms $\dot y_i \dot y_j$  \citep{2001JGR...10627933W}. This approximation may be adequate for the Moon with its small amplitudes of libration but is not applicable for a wide range of triaxial bodies including asteroids and minor planets. It is therefore of interest to compare these analytical estimates with the accurately computed frequency spectra of integrated trajectories.

Using the same simulation described in Sect. \ref{int.sec} and presented in Fig. \ref{pole.fig}, we apply the standard periodogram analysis to the solved functions of time representing the angular velocity and orientation of the model object. Our technical implementation of periodogram analysis is based on a direct least-squares fit of a constant term and $\sin$- and $\cos$- harmonics (three fitting terms per test frequency) on discretized trajectories as functions of time. This computation is relatively slow but it is free of biases that arise when a Fourier transform is used on truncated time series. The amplitude spectrum is computed from the estimated coefficients of the trigonometric terms. The duration of each integration is 200 orbits, and the sampling step of the output functions is $P_{\rm orb}/10$. Each periodogram least-squares solution is computed for a dense grid of 3000 test periods, which is equally spaced on the logarithmic scale, allowing us to detect and resolve even the sharpest spectral peaks in the high-frequency part of the spectrum. Using these well-tested and verified techniques, we find that
unlike the 1D (planar) case, the velocity vector variations are not equivalent in the body frame and in the inertial world frame. The former ($\boldsymbol \omega$) is related to physics, while the latter ($\boldsymbol \Omega$) is what an external observer can measure. 

\begin{figure*}
\centering
\includegraphics[width=.42\textwidth]{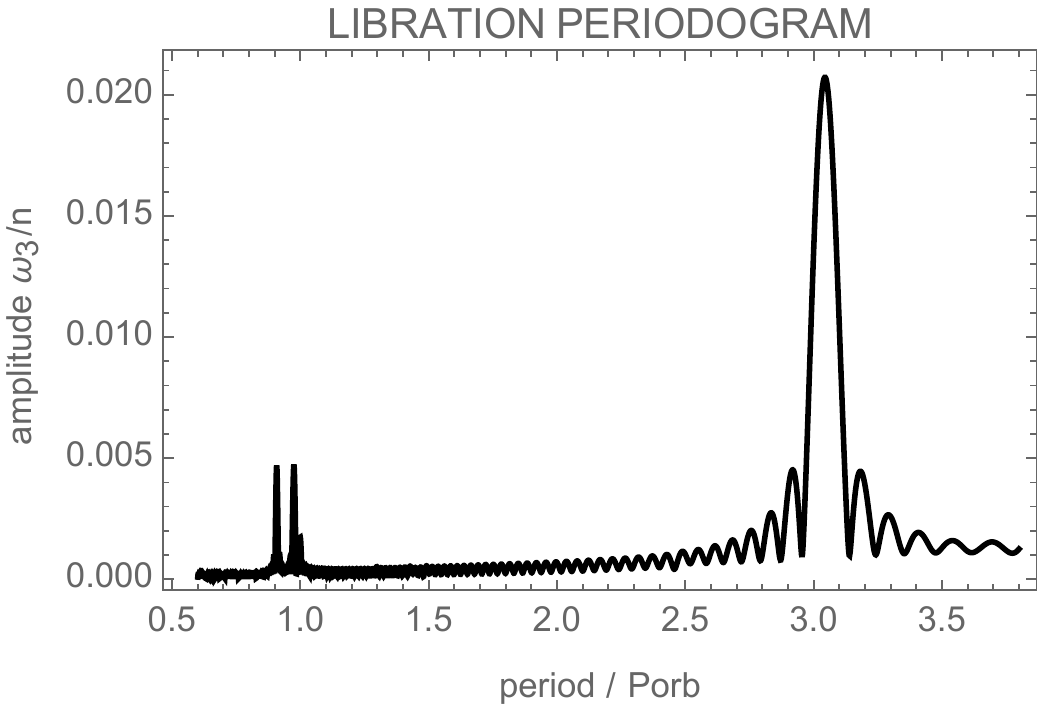}
\includegraphics[width=.42\textwidth]{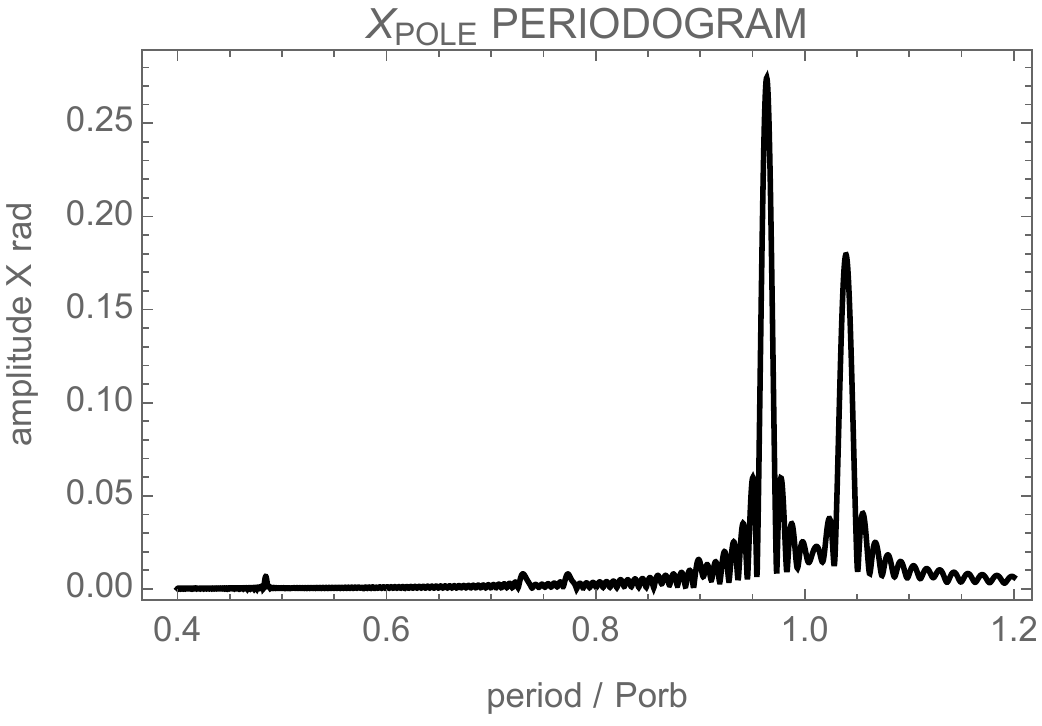}
\caption{Periodograms of the longitudinal angular velocity component $\omega_3$ in the body frame (left) and the $X_1$ coordinate of the north pole in the orbit plane for a test body with
$L_f=1$, $e=0.0047$, $n=4.5855$ d$^{-1}$ (model Enceladus) and initial
conditions described in the text. \label{perio.fig}}
\end{figure*}

Fig. \ref{perio.fig}, left, shows a short-period portion of the amplitude spectrum of the longitudinal component $\omega_3$ of angular velocity normalized to the mean motion $n$. In the harmonic oscillator approximation \citep{2011CeMDA.109...85R}, the period of the free libration mode is $P/P_{\rm orb}=(3\,(B-A)/C)^{-\frac{1}{2}}=3.054$ with the parameters of our Enceladus model. The actual location of the highest peak in the periodogram is 3.046. This small difference can be attributed to the deviation of our numerical solution from the idealized harmonic oscillator approximation. The dominating mode (not shown in the plot) by far, however, has a much longer period of $P/P_{\rm orb}=15.67$. This mode makes the velocity vector projected onto the equator describe a closed elliptical loop in Fig. \ref{pole.fig}, left. A more surprising outcome is that the forced libration mode is split between two frequencies that are both shifted with respect to the mean motion. These two split modes of nearly equal amplitude have normalized periods of $\sim 0.91$ and $0.975$. The only expected signal from the 1D analysis at exactly $P_{\rm orb}$ is also present but with a much smaller amplitude. A similar periodogram for the sidereal component $\Omega_3$ also shows a double peak presumably caused by the variable torque at  $P/P_{\rm orb}=0.964$ and 1.039 but no trace of a free libration mode with a period around 3. The greatest libration mode in $\Omega_3$ has a relative period of 15.67 (identical to $\omega_3$). Finally, the $X_1$ coordinate of the north pole projected onto the plane of orbit, which is a measure of the variable obliquity, shows a double peak in \ref{perio.fig}, right, at the same eigenfrequencies as $\Omega_3$, as expected, but no sign of a free libration mode between this feature and the 15.67 mode. The periodogram of $X_2$ is identical with the same modes of libration shifted in phase. The analytical expressions provided by \citet{2011CeMDA.109...85R} for libration in latitude do not match well these results. We interpret these differences and additional features in the spectra of 3D rotation as nonlinear effects of the initial roll and pitch velocity components via the free rotation terms that have been neglected in the theoretical studies.

\section{Mapping the vast chaotic swamp}
\label{cha.sec}
We now depart from our basic Enceladus model with $e=0.0047$ and $L_f=1$ explored in Sect. \ref{int.sec} and \ref{fre.sec} and investigate the rotation behaviour of test bodies with different orbital eccentricities and degrees of prolateness. The GALI index (Sect. \ref{gali.sec}) with $k=2$ and $t_{\rm max}=275$ orbits allows us to map this behaviour, i.e., represent it as an image in the corresponding 2D parameter space. However, the results also depend on the initial conditions of integration. Most of the close-in planets and moons rotate relatively slowly being in spin-orbit resonance, while the Solar system asteroids seem to have fast rotation rates outside resonance. We therefore pay special attention to the close vicinity of low-order spin-orbit resonances 1:1 (the Moon, Titan, Enceladus, etc.) and 3:2 (Mercury). 

\subsection{Synchronous rotation or 1:1 spin-orbit resonance}
In our first large-scale experiment, we fix the initial conditions at $z_1=z_2=z_3=0$ (see the definition of the attitude angles in Appendix A), $\dot y_1=\dot y_2=0$, and $\dot y_3=1.04\,n$. This initial point is close to the perfect 1:1 resonance, favouring order in its contest with chaos. The results are relevant for most of the moons in the Solar system, which are synchronized and have small amplitudes of free libration. Using the faster quaternion technique (Appendix B), we integrated the 7 rotation parameters for 275 orbits for a 2D grid of 113 values of eccentricity in the interval $[0,0.99]$ and 167 values of $L_f$ in the interval $[0.5,5.0]$, thus, $18,871$ integrations in total. The corresponding ratios of inertia moments $(B-A)/C$, $(C-A)/B$, and $(C-B)/A$
can be calculated using formulae in Sect. \ref{ecc.sec}. 

\begin{figure}
  \centering
\includegraphics[width=.95\columnwidth]{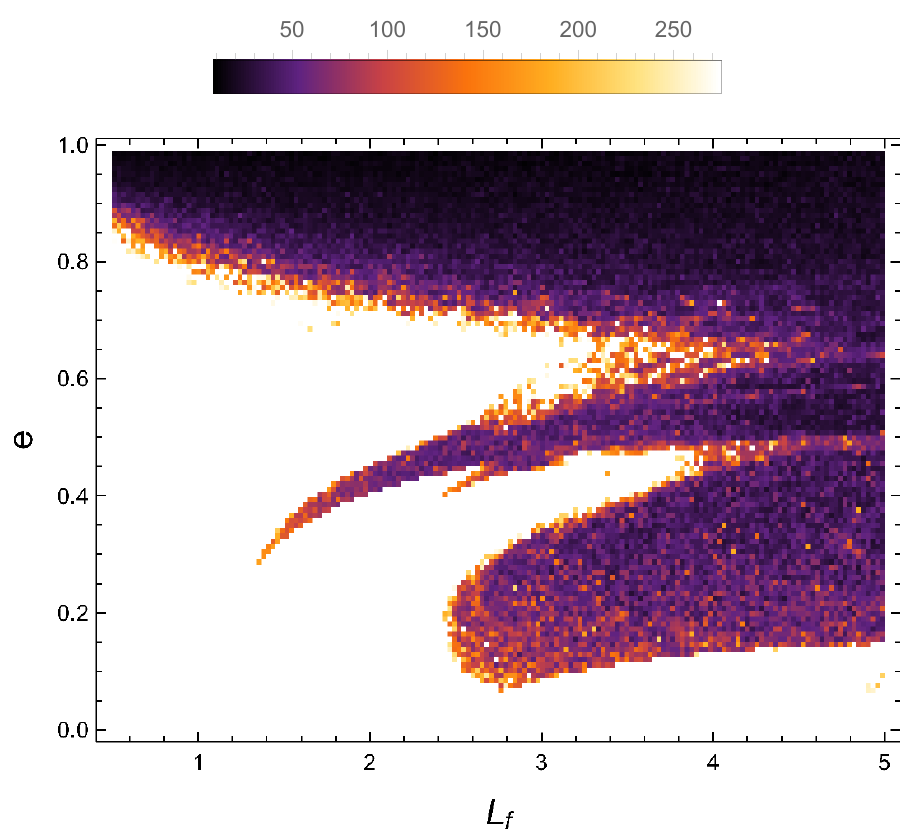}
\caption{Map of rotation states in the cross-section of the shape parameter $L_f$ and orbital
eccentricity $e$ for the close vicinity of the 1:1 spin-orbit resonance, $\omega_3(0)=1.04\, n$. The GALI$(2)$ index (see text) is color-coded in such a way that non-chaotic
rotation states are white and chaotic rotation states are dark-colored. Each GALI trial includes 275 orbits. \label{stormcloud.fig}}
\end{figure}

Fig. \ref{stormcloud.fig} shows the resulting distribution of the number of orbits $m_{\rm GALI}$ required for the GALI(2) index to drop below a threshold value of $10^{-12}$. The plot is color-coded such that the white color corresponds to the maximum number of orbits 275. If the GALI index does not reach the threshold value in 275 orbits, the corresponding trajectory may be regular. If, on the other hand, it drops below the threshold value in $m_{\rm GALI}$ orbits, the trajectory is certainly chaotic. The smaller this number, the larger is the exponent of chaotic divergence. Therefore, the darker colors show the trajectories of faster chaos.
Prior to performing this experiment, we expected to see the prevalence of chaotic rotation for high eccentricity and more extended shapes. We now see unexpected structures in the map such as the extended feature at mid-$e$, called ``the scimitar" in the following, with a tip at approximately $L_f=1.33$, $e=0.29$. An extensive zone of chaos at lower eccentricity and highly elongated shape is also noted.

\subsection{Orbital eccentricity destroys stability}
\label{ore.sec}
\begin{figure*}
\centering
\centerline{{\Large $e=0.10$} \ \ \ \ \ \ \ \ \ \ \ \ \ \ \ \ \ \ \ \ \ \ \ \ \ \ \ \ \ \ \ \ \ \ \ \ \ \ \ \ \ \ \ \ \ \ \ \ \ \ {\Large $e=0.65$}}
\includegraphics[width=.48\textwidth]{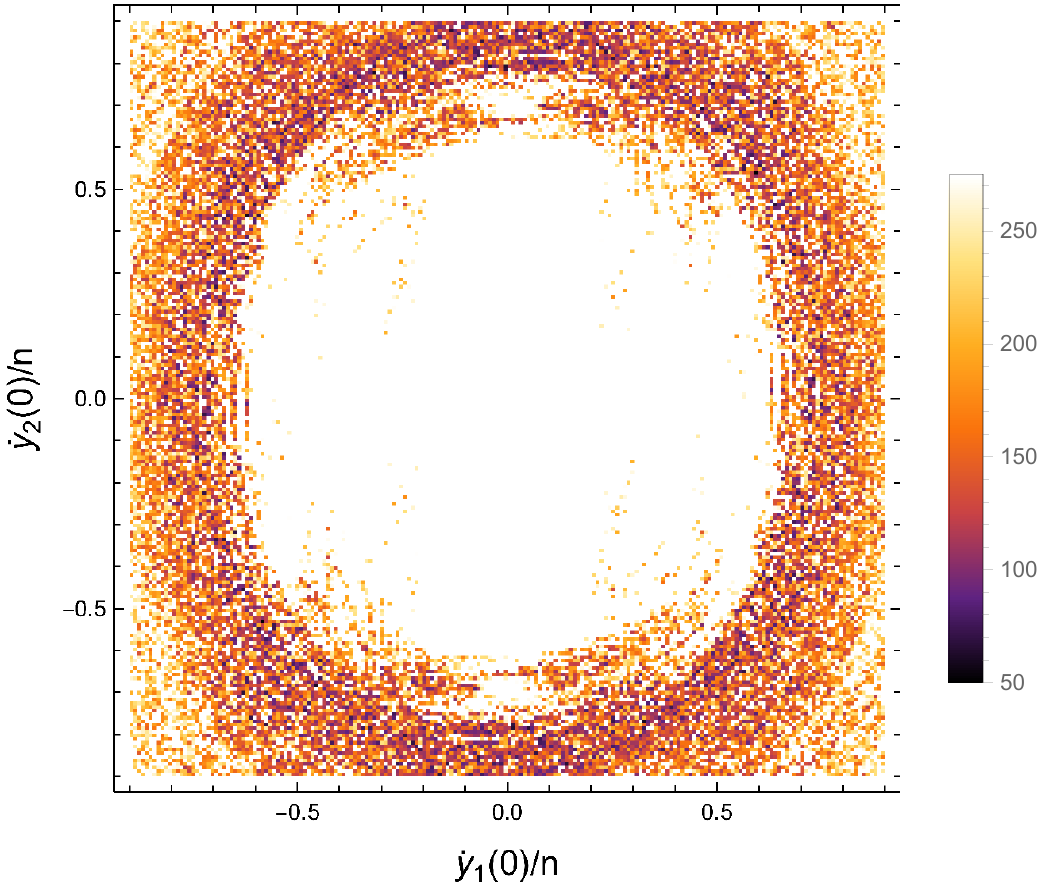}\quad
\includegraphics[width=.48\textwidth]{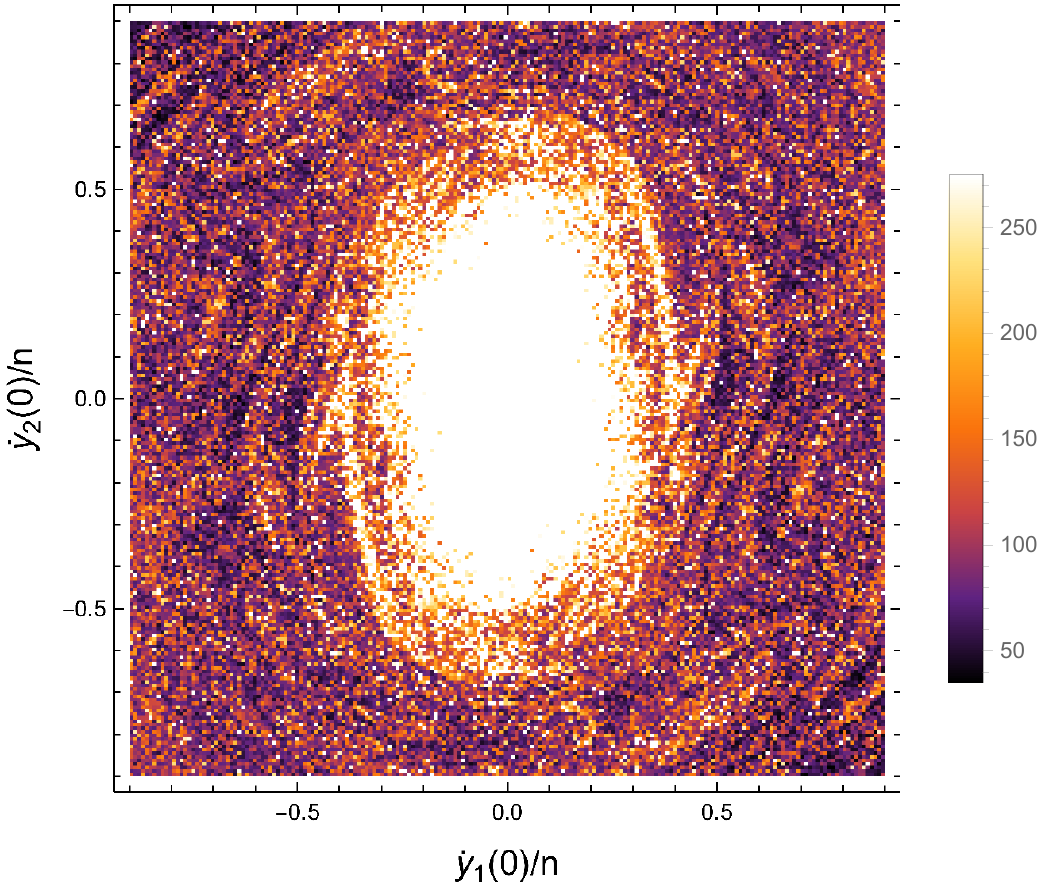}
\medskip
\centerline{{\Large $e=0.75$}\ \ \ \ \ \ \ \ \ \ \ \ \ \ \ \ \ \ \ \ \ \ \ \ \ \ \ \ \ \ \ \ \ \ \ \ \ \ \ \ \ \ \ \ \ \ \ \ \ {\Large $e=0.85$}}
\includegraphics[width=.48\textwidth]{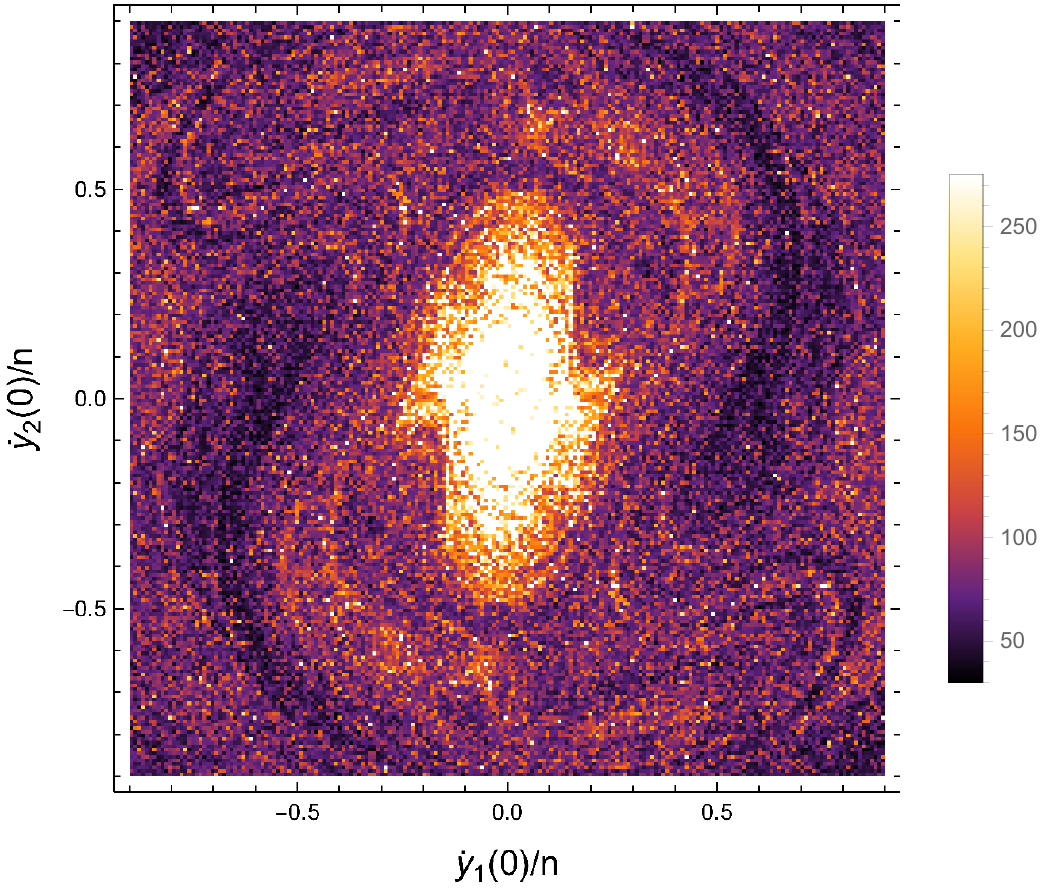}\quad
\includegraphics[width=.48\textwidth]{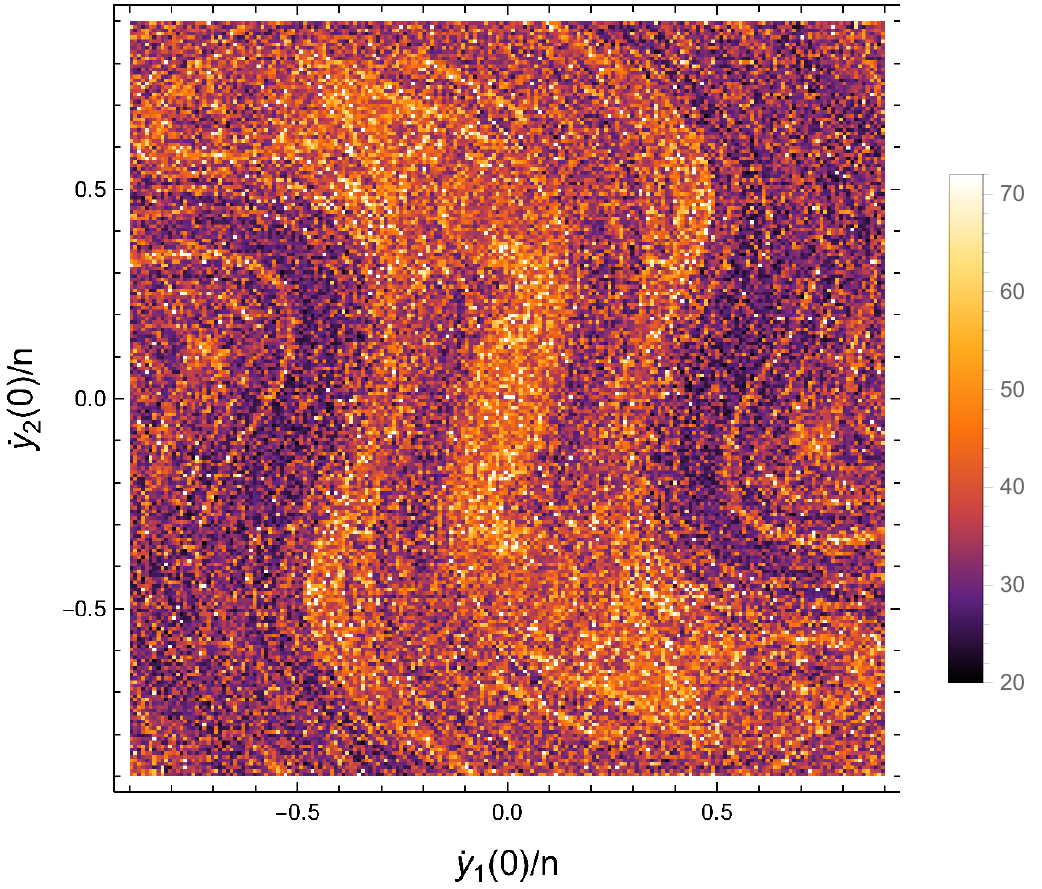}
\caption{Map of rotation states in the cross-section of initial
rotation velocity perturbations $\dot y_1(0)$ and
$\dot y_2(0)$, with initially synchronous rotation $\dot y_3(0)=1\,n$. The GALI$(2)$ index (see text) is color-coded in such a way that non-chaotic
rotation states are white and chaotic rotation states are dark-colored. Upper left: $e=0.1$; upper right: $e=0.65$; lower left: $e=0.75$; lower right: $e=0.85$. The base model moments of inertia ratios are $(B-A)/C=0.03573$, $(C-A)/B=0.04669$, and $(C-B)/A=0.01097$.\label{e00.fig}}
\end{figure*}

Fig. \ref{stormcloud.fig} suggests that chaotic rotation takes over for the base Enceladus model with $L_f=1$ approximately at $e=0.8$. What happens with the zone of regular rotation at a smaller $e$? Is Enceladus' rotation always stable or it can become chaotic with sufficiently strong perturbations from the nearly perfect synchronization? To answer these questions, we performed another series of large-scale numerical experiments keeping $L_f=1$ for a $197\times 213$ grid of initial velocities $\dot y_1$ (roll) and $\dot y_2$ (pitch) covering the interval $[-0.9,+0.9]\,n$ and a fixed $\dot y_3=1.04\,n$. Each of the $41,961$ integrations was performed for several values of eccentricity, included 275 orbits and yielded a $m_{\rm GALI}$ index. The results are presented in Fig. \ref{e00.fig} as color-coded maps with the same color scheme for comparison.

The progression of chaoticity maps for $e=0.10$, 0,65, 0.75, and 0.85 with the same other parameters clearly illustrates how the island of stability around the point of dynamical equilibrium corresponding to the synchronous spin-orbit resonance shrinks with increasing eccentricity and completely vanishes  between 0.75 and 0.85. There is no distinct boundary between chaotic and regular zones; in fact, small patches of chaotic values make incursions into the zone of stable rotation on all sides making a possibly fractal pattern. The chaotic zone is not uniform either with a fine structure indicating rapidly changing characteristic exponents. The maps are axially symmetric indicating that the chaoticity index is an even function of initial velocity perturbation.

\subsection{The crumbling islands of stability}
\begin{figure}
\centering
\includegraphics[width=.44\textwidth]{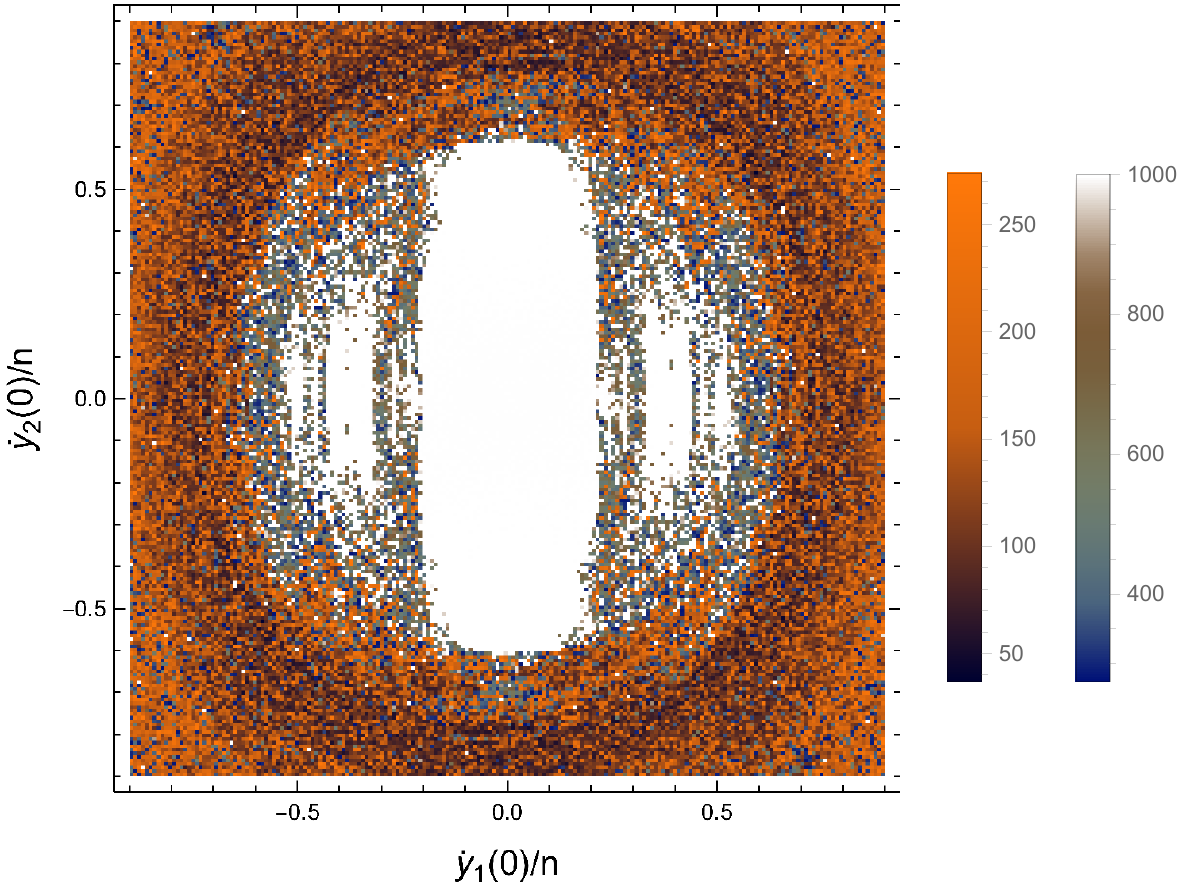}\quad
\caption{Map of rotation states in the cross-section of initial
rotation velocity perturbations $\dot y_1(0)$ and
$\dot y_2(0)$ for $e=0.1$. The GALI$(2)$ index (see text) is color-coded in such a way that non-chaotic
rotation states are white and chaotic rotation states are colored. The model parameters are the same as in Fig. \ref{e00.fig} for $e=0.1$ but 1000 orbits have been integrated instead of 275. The split color scheme is used to discriminate between the fast chaotic states achieved within 275 orbits (rust tones) and the slower chaotic states achieved between 275 and 1000 orbits (terrain tones). \label{island.fig}}
\end{figure}

The white areas in Figs. \ref{stormcloud.fig} and \ref{e00.fig} indicate that corresponding rotational trajectories may be regular and stable. The shredded appearance of the boundary between possible order and definite chaos suggests that some of these trajectories are also chaotic and simply did not have enough time to reach the threshold GALI value. To test this hypothesis, we performed the same simulation as shown in Fig. \ref{e00.fig} for $e=0.1$ but with 1000 orbits for each grid point instead of 275. The result is shown in Fig. \ref{island.fig}.

\begin{figure*}
\centering
\includegraphics[width=.51\textwidth]{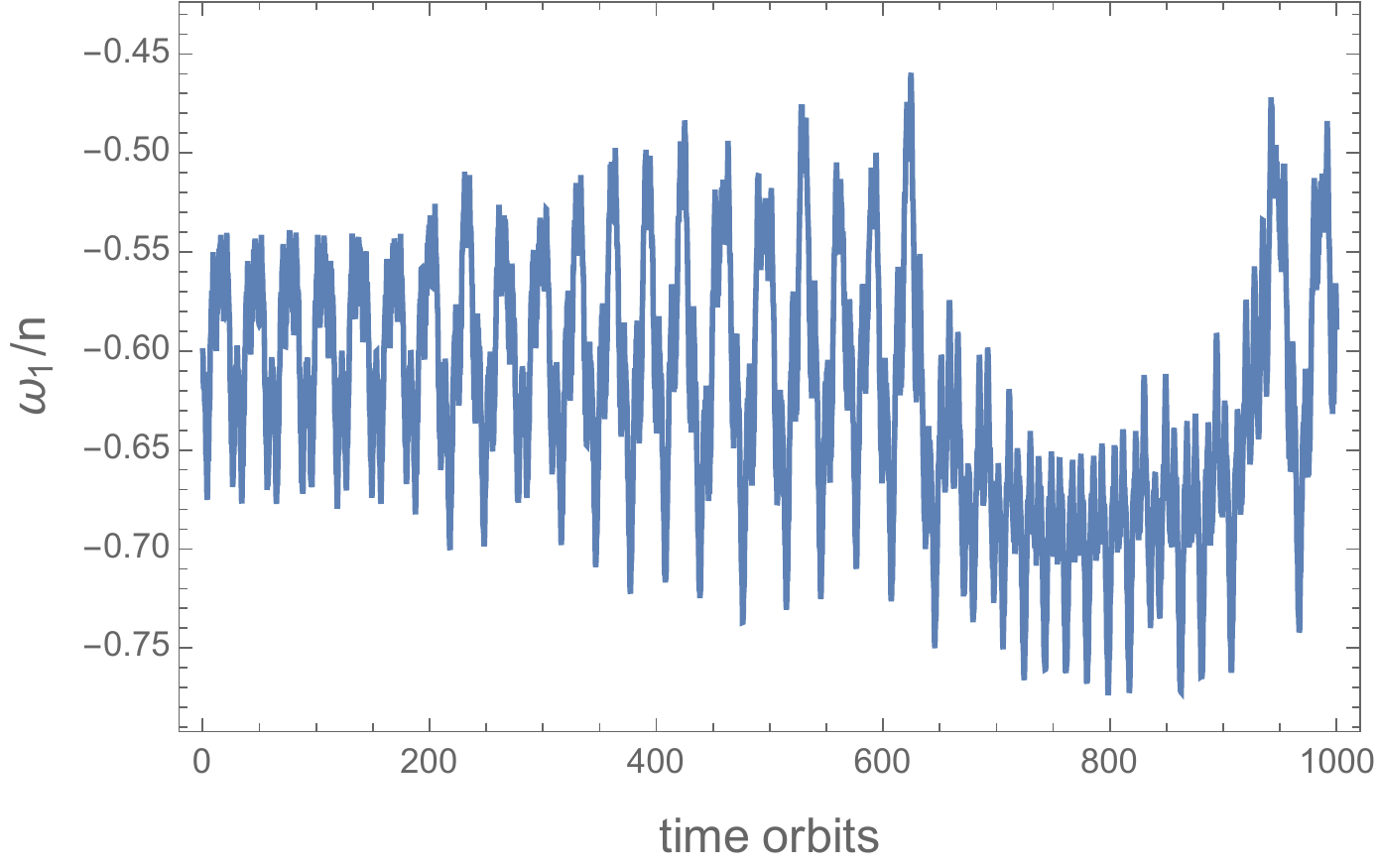}
\includegraphics[width=.34\textwidth]{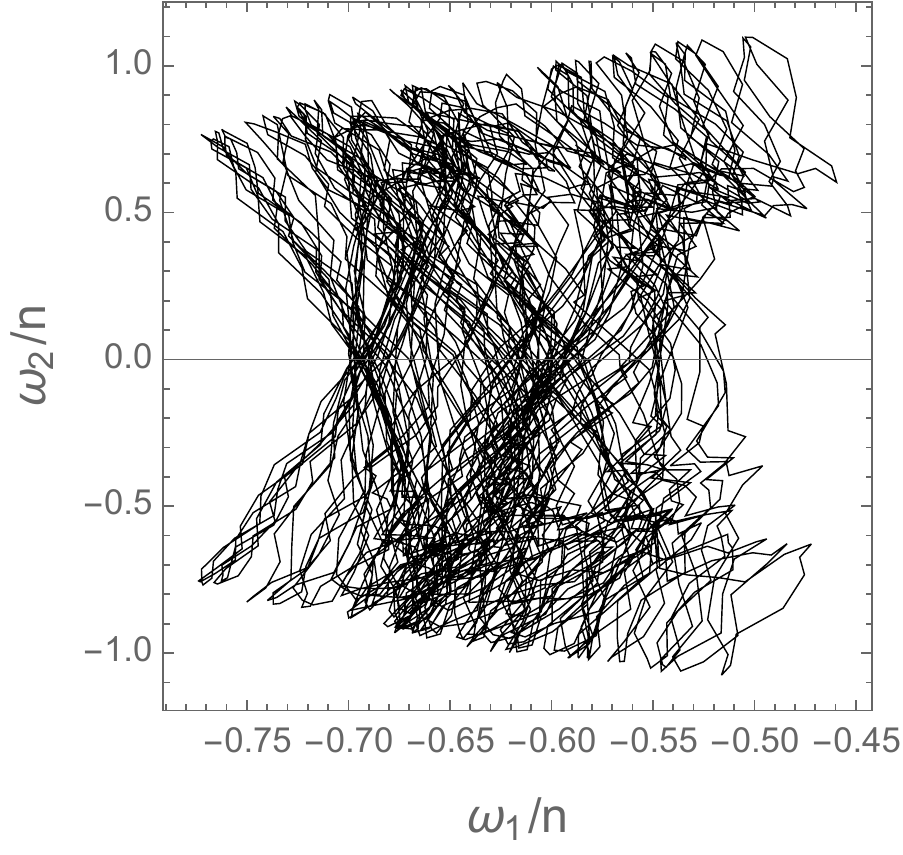}
\caption{Trajectories of integrated rotation of the test body (the same as in Fig. 1) with $e=0.1$, initial orientation perfectly aligned with the direction to the primary at pericenter, and initial velocities $\dot y_1(0)=-0.6\, n$,  $\dot y_2(0)=0$, $\dot y_1(0)=1.0\, n$, representing an initial point with a GALI(2) index between 275 and 1000 in Fig. \ref{island.fig}. The roll (left plot) and pitch (right plot) components of spin rate are shown as a function of time for 1000 rotations and a parametric plot, respectively.  \label{traj.fig}}
\end{figure*}

Chaos has made considerable advances with longer integration times and encroached the large island of stability centered on the point of perfect 1:1 resonance on all sides, but more so in the $\dot y_1(0)$ dimension (roll angle) where the restoring gravitational torque is the weakest. The map has become even more shredded and finely structured. The white spaces may still not represent non-chaotic trajectories because 1000 orbits is just an instant on the scale of astronomical lifetimes.  Fig. \ref{traj.fig} shows an example of integrated trajectories with initial parameters in the domain where GALI(2) index acquires values above 275 but below 1000, namely, the initial attitude quaternion at pericenter $q(0)=(1,0,0,0)$, initial velocity $\dot{\boldsymbol y}(0)=(-0.6,0,1)\, n$, and $e=0.1$ for the same model body. The spin rate component as a function of time (left panel) may seem to be confined and quasi-periodic for a few hundred orbits but becomes obviously chaotic toward the end of this simulation. The parametric plot of the spin rate trajectory (right panel) is drastically different from orderly precession patterns seen in Fig. 1, left panel.

\subsection{3:2 spin-orbit resonance}

The 1:1 spin-orbit resonance (synchronous rotation) is the main stable state for planetary satellites in the Solar system, which mostly have small eccentricities and are sufficiently close to to their hosts. The Moon, however, could be easily captured into a 3:2 spin-orbit resonance if its initial rotation was prograde and fast \citep{2013MNRAS.434L..21M}. Probabilities of capture into higher spin-orbit resonances can be semi-analytically solved for in the 1D case with simplified toy tidal models such as the constant phase lag model and with more advanced nonlinear models \citep{gol, 1968ARA&A...6..287G, 2012ApJ...752...73M}. Mercury is currently in the 3:2 resonance but the outcome could have been different depending on the past chaotic evolution of its eccentricity \citep{noy}. Given that asteroids and minor planets often have higher eccentricities, it is of interest to see how the domain of chaotic rotational states looks like for supersynchronous resonances.

\begin{figure}
  \centering
\includegraphics[width=.44\textwidth]{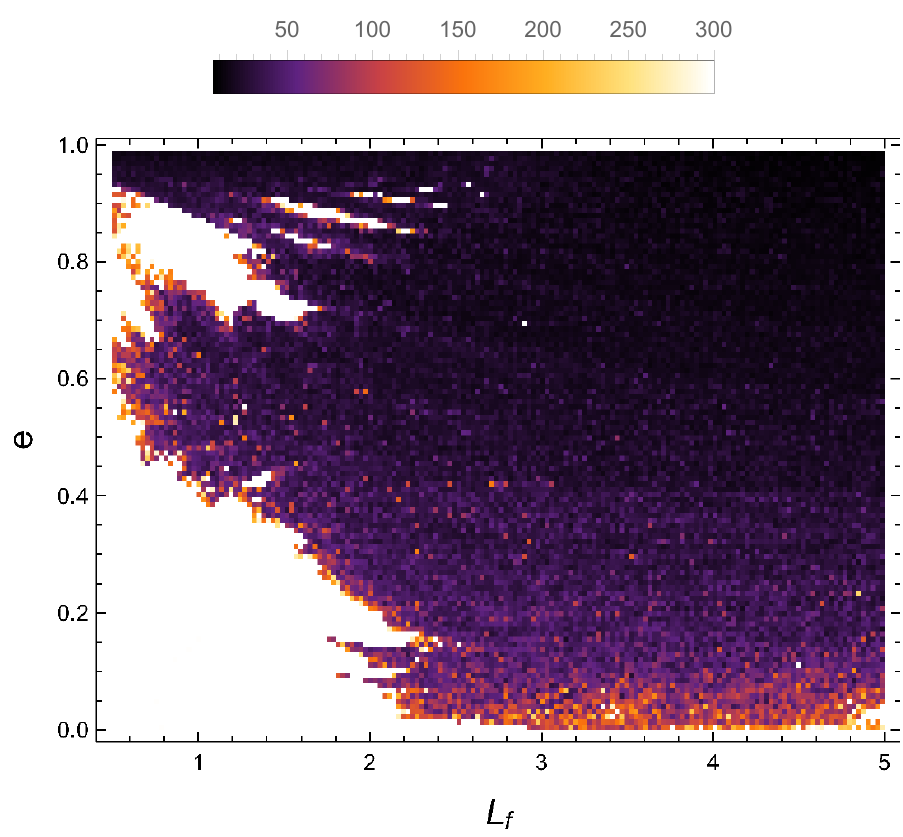}\quad
\caption{Map of rotation states in the cross-section of the shape parameter $L_f$ and orbital
eccentricity $e$ for the close vicinity of the 3:2 spin-orbit resonance, $\omega_3(0)=1.54\, n$. The GALI$(2)$ index (see text) is color-coded in such a way that non-chaotic
rotation states are white and chaotic rotation states are dark-colored. Each GALI trial includes 300 orbits. \label{storm15.fig}}
\end{figure}

We performed massive integrations for the base Enceladus model (Sect. \ref{cha.sec}) with fixed initial conditions $z_1=z_2=z_3=0$ , $\dot y_1=\dot y_2=0$, and $\dot y_3=1.54\,n$. This initial point is close to the perfect 3:2 resonance, with a small deviation meant to introduce a low-amplitude longitudinal libration. Using the faster quaternion technique (Appendix B), we integrated the 7 rotation parameters for 300 orbits for a 2D grid of 113 values of eccentricity in the interval $[0,0.99]$ and 167 values of $L_f$ in the interval $[0.5,5.0]$, thus, $18,871$ integrations in total. Each integration computed the GALI(2) value for 300 orbits or until it dropped below the threshold. The resulting numbers of orbits are shown in Fig. \ref{storm15.fig} as a color-coded map. 

Compared to the analogous computation for the 1:1 resonance shown in Fig. \ref{stormcloud.fig}, we note a much wider spread of chaos, which made incursions into the domain of quasi-stable states at low eccentricity and small prolateness values. Rotation in the vicinity of $1.5\,n$ becomes strongly chaotic for $L_f=1$ already at $e\simeq 0.4$. Surprisingly, there is another area of stable rotation for $e>0.75$, as well as a few smaller islands of stability for more elongated bodies at higher eccentricities. The scimitar feature in Fig. \ref{stormcloud.fig} is completely gone, instead, a few smaller streaks are present at low eccentricity around $L_f=2$.

\begin{figure*}
\centering
\centerline{{\Large $e=0.3$} \ \ \ \ \ \ \ \ \ \ \ \ \ \ \ \ \ \ \ \ \ \ \ \ \ \ \ \ \ \ \ \ \ \ \ \ \ \ \ \ \ \ \ \ \ \ {\Large $e=0.4$}}
\includegraphics[width=.48\textwidth]{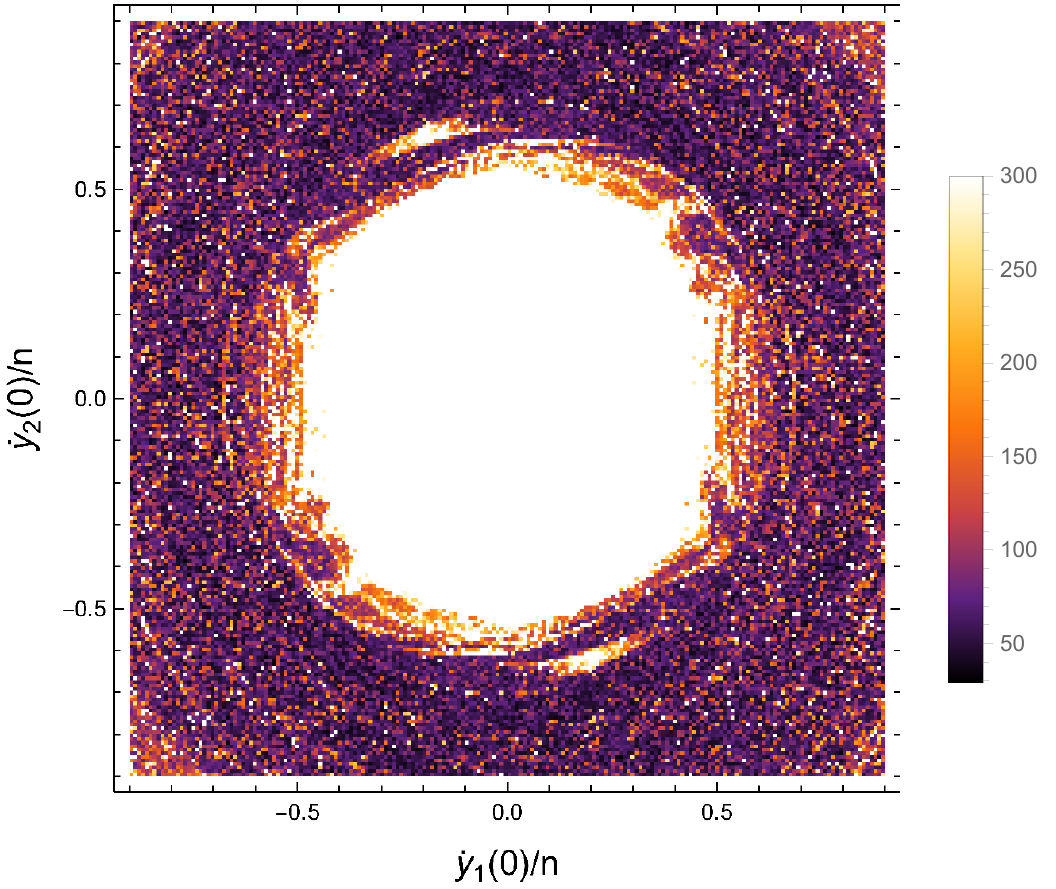}\quad
\includegraphics[width=.48\textwidth]{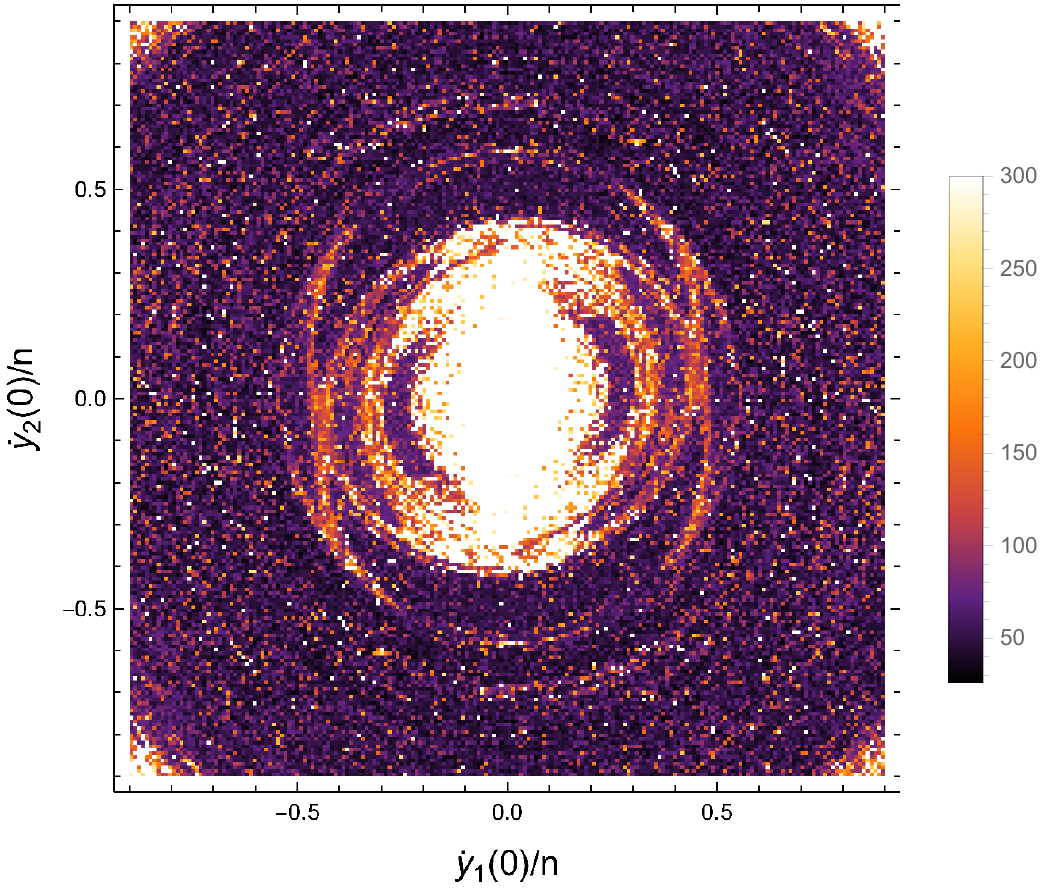}
\caption{Map of rotation states in the cross-section of initial
rotation velocity perturbations $\dot y_1(0)$ and
$\dot y_2(0)$, with initially prograde rotation in the 3:2 resonance $\dot y_3(0)=1.5\,n$. The GALI$(2)$ index (see text) is color-coded in such a way that non-chaotic
rotation states are white and chaotic rotation states are dark-colored. Left: $e=0.3$; right: $e=0.4$. The base model moments of inertia ratios are $(B-A)/C=0.03573$, $(C-A)/B=0.04669$, and $(C-B)/A=0.01097$.\label{e15.fig}}
\end{figure*}

This numerical experiment shows that there are islands of stable rotation (at least, within 300 orbits) surrounding the 3:2 resonance. How large are they? Do they also shrink with increasing eccentricity? To answer these questions, we performed numerical simulations for the base Enceladus model with $L_f=1$ on a grid of $e$ analogous to the integrations described in Sect. \ref{ore.sec}.
For a $197\times 213$ grid of initial velocities $\dot y_1$ (roll) and $\dot y_2$ (pitch) covering the interval $[-0.9,+0.9]\,n$ and a fixed $\dot y_3=1.5\,n$, up to 300 orbits orbits were computed yielding a $m_{\rm GALI}$ index. The results are shown in Fig. \ref{e15.fig} as color-coded maps with the same color scheme for comparison. A wide stability island is found at $e=0.3$. It rapidly dwindles as eccentricity increases toward $e=0.4$, and completely disappears at slightly larger $e$. There is no area of stable rotation at $e=0.5$ or $e=0.7$. However, a similar map for $e=0.8$ (not reproduced for brevity) shows again a sizeable island of stability centered on the 3:2 resonance. The boundaries are complex and shredded indicating sharp variations of maximum Lyapunov exponents with initial conditions.

\section{Summary}
\label{sum.sec}
Rotational dynamics of elongated celestial bodies in two-body systems is a battlefield of chaos versus order. Chaos clearly prevails across the parameter space outside of the domain of low eccentricity and nearly spherical shapes. The maps of chaoticity in the eccentricity--shape space as measured by the $m_{\rm GALI}$ index show complex structures and shredded boundaries in the vicinity of the lowest order resonances 1:1 and 3:2. The islands of stable rotation centered on these resonances tend to shrink with increasing eccentricity and completely vanish at some limiting values. Even quasi-stable trajectories in the vicinity of perfect resonances may be in reality chaotic with longer characteristic times than the duration of integrations implemented in this paper. We find that chaotic trajectories of highly variable maximum exponents are tightly packed in the space of initial parameters. Isolated areas of quasi-stable rotation unexpectedly appear for the vicinity of the 3:2 spin-orbit resonance at high values of eccentricity.

The commonly used simplification of the Euler's equations of motion by omitting the inertial terms and approximating the ODE with a harmonic pendulum does not give a full and accurate account of the spectrum of free libration, nutation, and polar motion for the 1:1 resonance. We find that the forced longitudinal libration mode splits into two frequencies that are either shifted away from the orbital frequency or bracket it depending on which coordinate frame is used. The dominating mode of libration in all three angles of orientation and velocity components is not predicted by theory. It corresponds to a relatively slow precession-like variation of obliquity. We could not find the classical free libration mode in longitude for velocity components in the world reference frame, i.e., as seen by an external observer. Rotational motion in the vicinity of higher spin-orbit resonances is more complex, governed by a set of non-commensurate eigenfrequencies.

This study opens up quite a few issues that have yet to be answered. One would like to know if there are islands of stable rotation at nonzero initial obliquity in pericenter, analogous to the Cassini state of the Moon. The observed distribution of rotation periods of main belt asteroids and their orbital eccentricities suggest that many of them may be in chaotic states. It remains to be seen if long-term stable resonances exist at high eccentricities and high rates of rotation. We would like to understand the pattern of coupled 3D motion that results in significantly different libration frequencies in the body and world reference frames. Tidal friction is likely a great regularizer of chaos for planetary satellites and close exoplanets, whereas its role for more distant and typically more eccentric minor planets is not well known. 

\section*{Acknowledgments}
DV gratefully acknowledges the support of the STFC via an Ernest Rutherford Fellowship (grant ST/P003850/1).

\section*{Data Availability}
The simulation inputs and results discussed in this paper are available upon reasonable request to the corresponding author.

\section*{Appendix A: Attitude solution with Euler angles}
Among a few mathematical representations of coordinate frame rotation,
we choose the sequence of elementary Euler's rotations 3-2-1, which is
standard in marine applications and space engineering \citep[for a   detailed derivation, see][]{fos}.
For any point
within the body with a central position vector $\hat{\boldsymbol r}$ in
the orbital frame, the position vector $\boldsymbol r$ in the body frame is obtained
through a sequence of right-handed rotations around axis 3, then around
axis 2, and, finally, axis 1:
\eb
\boldsymbol r=\boldsymbol R_1(z_1)\,\boldsymbol R_2(z_2)\,\boldsymbol R_3(z_3)\,\hat {\boldsymbol r}=\boldsymbol R_{321}(z_1,z_2,z_3)\,\hat{ \boldsymbol r}.
\label{hatr.eq}
\ee
The angles $z_i$, $i=1,2,3$, define the instantaneous orientation of the
body frame with respect to the orbital frame, hence, they are continuous functions of time. The elementary rotation matrices are
\begin{eqnarray}
\boldsymbol R_1 & = & \left[\begin{array}{ccc}
1 & 0 & 0 \\
0 & \cos{z_1} & \sin{z_1} \\
0 & -\sin{z_1} & \cos{z_1}
\end{array} \right] \nonumber \\
\boldsymbol R_2 & = & \left[\begin{array}{ccc}
\cos{z_2} & 0 & -\sin{z_2} \\
0 & 1 & 0 \\
\sin{z_2} & 0 & \cos{z_2}
\end{array} \right] \nonumber \\
\boldsymbol R_3 & = & \left[\begin{array}{ccc}
\cos{z_3} & \sin{z_3} & 0 \\
-\sin{z_3} & \cos{z_3} & 0 \\
0 & 0 & 1
\end{array} \right] 
\end{eqnarray}
The matrices $\boldsymbol{R}_i$ are orthogonal, as well as their
product $\boldsymbol{R}_{321}$, and $\boldsymbol{R}_i^{-1}(z_i)=\boldsymbol{R}_i^{T}(z_i)=\boldsymbol{R}_i(-z_i)$. The order of elementary rotations is important because the same
orientation of the body frame can be represented by another sequence of
elementary rotations through different angles.

As we noted before, there is no direct mapping between the angles
$\{z_1,z_2,z_3\}$ and $\{y_1,y_2,y_3\}$. Their time derivatives, on the
other hand, are related by the same sequence of geometrical rotations
and projections onto the basis vectors of the rotated frames. This is
true because of the invariance of the rotation vector to a frame rotation,
\eb
\hat{\dot{\boldsymbol y}}=\boldsymbol R_{321}(z_1,z_2,z_3)\, \dot{\boldsymbol y},
\ee
which is a direct consequence of the invariance of the vector cross product to unitary transformations, $\hat{\boldsymbol\Omega}= \hat{\boldsymbol\omega}\times \hat{\boldsymbol r}=(\boldsymbol R_{321}(z_1,z_2,z_3)\,\boldsymbol\omega) \times(\boldsymbol R_{321}(z_1,z_2,z_3)\,\boldsymbol r)=\boldsymbol R_{321}(z_1,z_2,z_3)\,
(\boldsymbol\omega \times \boldsymbol r)$. Considering that the angular
rates $\{\dot z_1,\dot z_2,\dot z_3\}$ are the geometric projections
of the spin vector $\hat{\dot{\boldsymbol y}}$ onto the basis vectors of 
sequentially rotated frames (starting with the orbital frame), we can
write
\begin{eqnarray}
\dot z_1 & = & (\boldsymbol{R}_1^{-1}\,\dot{\boldsymbol y})[[1]] =  \dot{\boldsymbol y}[[1]] \nonumber \\
\dot z_2 & = & (\boldsymbol{R}_2^{-1}\,\boldsymbol{R}_1^{-1}\,\dot{\boldsymbol y})[[2]] = (\boldsymbol{R}_1^{-1}\,\dot{\boldsymbol y})[[2]]\nonumber \\
\dot z_3 & = & (\boldsymbol{R}_3^{-1}\,\boldsymbol{R}_2^{-1}\,\boldsymbol{R}_1^{-1}\,\dot{\boldsymbol y})[[3]] = (\boldsymbol{R}_2^{-1}\,\boldsymbol{R}_1^{-1}\,\dot{\boldsymbol y})[[3]],
\label{dotz.eq}
\end{eqnarray}
where $[[i]]$ denotes $i$th component of a 3-vector. 

The same transformation in the reverse order can be written as \citep
[cf. ][Eq. 2.27]{fos}
\eb
[\dot y_1,\dot y_2,\dot y_3]^T = [\dot z_1,0,0]^T+\boldsymbol{R}_1\,[0,\dot z_2,0]^T+\boldsymbol{R}_1\,\boldsymbol{R}_2\,[0,0,\dot z_3]^T
\label{ome.eq}
\ee
which formally leads to a linear matrix equation in the form
\eb
[\dot y_1,\dot y_2,\dot y_3]^T = \boldsymbol T\,[\dot z_1,\dot z_2,\dot z_3]^T.
\label{t.eq}
\ee
We note that this is merely a formal representation of sequential geometric 
projections as the 3-vector $[\dot z_1,\dot z_2,\dot z_3]^T$ does not exist
in any 
specific coordinate frame. Eqs. \ref{t.eq} are further inverted arriving
at 
\eb
[\dot z_1,\dot z_2,\dot z_3]^T=\boldsymbol T^{-1}\,[\dot y_1,\dot y_2,\dot y_3]^T,
\ee
or, explicitly,
\begin{eqnarray}
\dot z_1 & = & \dot y_1 + \dot y_2\,\sin z_1\,\tan z_2+\dot y_3\,\cos z_1\,\tan z_2 \nonumber \\
\dot z_2 & = & \dot y_2\,\cos z_1-\dot y_3\,\sin z_1 \nonumber \\
\dot z_3 & = & \dot y_2\,\sin z_1\,\sec z_2+\dot y_3\,\cos z_1\,\sec z_2.
\label{z.eq}
\end{eqnarray}

The time derivatives of $z_i$ are burdened with an explicit singularity at $z_2=\pm \pi/2$. The reason for this singularity is that a second rotation by $\pi/2$ makes the rotated first axis to be aligned with the former third axis, which results in a degeneracy of the rotation schema. The same attitude can be obtained by changing $z_3$ and $z_1$ by an arbitrary angle in the opposite sense. Physically, this degeneracy is encountered when the pitch angle (related to
the obliquity of the equator in orbital dynamics) exceeds the critical value and the rotation becomes retrograde. It is noted that these time derivatives
in the inertial frame of the orbit refer to three mutually non-orthogonal directions, and the total spin
is therefore not conserved, $\dot z_1^2+\dot z_2^2+\dot z_3^2 \neq 
||\dot{\boldsymbol y}||^2$.

As a sanity check, we can derive the same equations for $\dot z_i$ as
in Eqs. \ref{dotz.eq}
considering the geometric projections of the spin vector $\hat{\dot{\boldsymbol y}}$ in the inertial orbital frame and performing
the elementary rotations in the direct 321 order. Indeed, it is readily
seen that $\dot z_3=\hat{\dot{\boldsymbol y}}[[3]]$, etc. Thus, the
chain of rotations is reversible maintaining the invariant spin vector.
Our approach implemented in this paper is to add Eqs. \ref{z.eq}
to the set of 2nd order ODEs of motion (\ref{eu.eq}).
The number of unknown functions is formally raised from 3 to 6, but
this hardly has any impact on the integration time. An additional benefit
is that the sidereal rotation velocity and the rotational axis nutation
are directly obtained from the same integration.

For the sake of completeness, we also provide an algebraic rigorous derivation
that does not require any geometric considerations. It involves a third coordinate
frame $\{\bar{\boldsymbol Y}_1, \bar{\boldsymbol Y}_2, \bar{\boldsymbol Y}_3\}$,
which is a rotating triad defined by the principal axes of inertia of the asteroid.
Any fixed point within the body has a constant position $\bar{\boldsymbol r}$
in this frame, and its velocity and acceleration are equal to zero. At any given time $t$ the orientation of this frame coincides with that of the previously
considered body frame $\{\boldsymbol Y_1,\boldsymbol Y_2,\boldsymbol Y_3\}$.
Therefore, from Eq. \ref{hatr.eq},
\eb
\hat{\boldsymbol r}=\boldsymbol R_3(-z_3)\,\boldsymbol R_2(-z_2)\,\boldsymbol R_1(-z_1)\,\bar {\boldsymbol r}.
\ee
Differentiating this equation with respect to time,
\eb
\begin{split}
\frac{d\hat{\boldsymbol r}}{dt}=\left(\dot{\boldsymbol R_3}(-z_3)\,\boldsymbol R_2(-z_2)\,\boldsymbol R_1(-z_1)+
\boldsymbol R_3(-z_3)\,\dot{\boldsymbol R_2}(-z_2)\,\boldsymbol R_1(-z_1)\right.\\+
\left.\boldsymbol R_3(-z_3)\,\boldsymbol R_2(-z_2)\,\dot{\boldsymbol R_1}(-z_1)\right)\,\bar {\boldsymbol r}.
\end{split}
\ee
where we use $\dot x$ for time derivative of $x$. By definition, $d\hat{\boldsymbol r}/dt=\hat{\boldsymbol\omega}\times \hat{\boldsymbol r}$, which can further
be equalled to $\boldsymbol R_3(-z_3)\,\boldsymbol R_2(-z_2)\,\boldsymbol R_1(-z_1)
(\boldsymbol\omega\times \boldsymbol r)$ due to the frame invariance of the spin vector. Note that we switched to the instantaneously inertial body frame
$\{\boldsymbol Y_1,\boldsymbol Y_2,\boldsymbol Y_3\}$. At any given time, 
$\boldsymbol r=\bar{\boldsymbol r}$, hence,
\eb 
\begin{split}
\boldsymbol\omega\times \boldsymbol r= \left(\boldsymbol R_1(z_1)\,\boldsymbol R_2(z_2)\,\boldsymbol R_3(z_3)\dot{\boldsymbol R_3}(-z_3)\,\boldsymbol R_2(-z_2)\,\boldsymbol R_1(-z_1)\right.\\+
\boldsymbol R_1(z_1)\,\boldsymbol R_2(z_2)\,\dot{\boldsymbol R_2}(-z_2)\,\boldsymbol R_1(-z_1)\\+
\left.\boldsymbol R_1(z_1)\,\dot{\boldsymbol R_1}(-z_1)\right)\,\boldsymbol r.
\end{split}
\ee
The cross product in the left-hand part of this equation can be replaced by
a scalar product $\boldsymbol S \,\boldsymbol r$, with $\boldsymbol S$ being
the corresponding skew-symmetric matrix. Using the three nonzero elements
above its diagonal and performing the matrix multiplication and summation
on the right-hand side obtains
\begin{eqnarray}
-\omega_1 &=& -\dot z_1 + \sin\,z_2\,\dot z_3 \nonumber \\
\omega_2  &=& \cos z_1\,\dot z_2 + \cos z_2\,\sin\,z_1\,\dot z_3 \nonumber \\
-\omega_3 &=& \sin z_1\,\dot z_2 - \cos z_1\,\cos\,z_2\,\dot z_3.
\end{eqnarray}
Solving this linear system of equations for $\{\dot z_1,\dot z_2,\dot z_3\}$,
one arrives at a solution equivalent to Eqs. \ref{z.eq}.

In principle, there is an alternative approach to solving the problem
of integrating the 3D equations of motion
by re-writing Euler's equations (\ref{eu.eq}) in terms of angles $z_i$
and their derivatives. \citet{1995Icar..117..149B} explored this route
to map the chaotic rotation of Hyperion. They used the more traditional
for celestial mechanics 3-1-3 schema of elementary Euler rotations to
represent Hyperion's instantaneous attitude, which is a matter of technical preference. After deriving a system of 1st order equations
for body frame spin rates analogous to Eq. \ref{ome.eq},
the vector of angular acceleration $[\ddot y_1,\ddot y_2,\ddot y_3]^T$
was computed by time-differentiating the right-hand part incorrectly
assuming that the rotation matrices $\boldsymbol{R}_i$ and their
arguments $z_i$ are independent physical entities. The resulting complex
trigonometric forms were inverted yielding 2nd order equations with
explicit singularities. The correct second derivatives $\ddot z_i$ can
be obtained by replacing all single-dotted terms in Eqs. \ref{dotz.eq}
with double-dotted variables. This follows from the fact that the angular
acceleration vector in the body frame is also rotation-invariant,
$\hat{\dot{\boldsymbol\Omega}}= \hat{\dot{\boldsymbol\omega}}\times \hat{\boldsymbol r}=(\boldsymbol R_{321}(z_1,z_2,z_3)\,\dot{ \boldsymbol\omega}) \times(\boldsymbol R_{321}(z_1,z_2,z_3)\,\boldsymbol r)=\boldsymbol R_{321}(z_1,z_2,z_3)\,
(\dot{\boldsymbol\omega} \times \boldsymbol r)$ for any fixed position 
$\boldsymbol r$. The second derivatives of the attitude angles depend
therefore only on the physical acceleration vector and the attitude angles
themselves but not on their first derivatives.

For an instantaneous unit vector $\boldsymbol{p}$ toward the perturbing
body (Saturn in the case of Enceladus), the direction cosine products
in Eqs. \ref{eu.eq} are the corresponding elements of the outer product
of $\boldsymbol{p}$ with itself:
\eb
s_i\,s_j=(\boldsymbol{p}\otimes \boldsymbol{p})[[i,j]].
\ee
This vector is computed from the known attitude and the direction vector
in the inertial orbital frame:
\eb
\boldsymbol p=\boldsymbol R_{321}(z_1,z_2,z_3)\,\hat{ \boldsymbol p},
\ee
The gravitating perturber always stays in the initial orbit plane, and
its position is defined by a single true anomaly angle $f(t)$:
\eb
\hat{ \boldsymbol p}=[\cos f,\sin f,0]^T.
\ee
The required trigonometric functions of true anomaly are computed from
\begin{eqnarray}
\cos f &=& \frac{\cos E -e}{1-e\cos E} \nonumber\\
\sin f &=& \frac{\sqrt{1-e^2}\sin E}{1-e\cos E},
\end{eqnarray}
where $e$ is eccentricity and $E$ is the eccentric anomaly as a function of time.
The latter orbital parameter is computed either by the back-interpolation scheme
of \citet{2018arXiv181202273T}
or by the optimized high-accuracy method by M. Murison of USNO\footnote{\url{http://murison.alpheratz.net/dynamics/twobody/KeplerIterations\_summary.pdf}} from the mean anomaly $M(t)=(t-t_0)n$.

The system of six ODEs includes three equations of second order. The required number
of initial conditions is 9. We chose to define the initial angles $y_1(0)=y_2(0)=y_3(0)=0$ (by definition) and the initial angular velocity in the body
frame $\boldsymbol \omega=[\dot y_1(0),\dot y_2(0),\dot y_3(0)]$, as well as initial Euler
angles of the attitude matrix, $z_1(0),z_2(0),z_3(0)$.

\section*{Appendix B: Attitude solution with quaternion representation}

A well-known alternative to the Euler rotation matrix representation of rigid
rotation is the use of attitude quaternions. It formally includes more time-variable functions but it is also numerically stable for any type of motion, free of singularities, and computationally faster because it does not involve as many trigonometric functions.

The orientation (attitude) of a rotating body is represented by a  unit quaternion $q = (q_0,q_1, q_2,q_3)$, where each of the components is a smooth
function of time. Alternatively, it can be written as $(q_0,\boldsymbol q)$ separating its scalar part and 3-vector part. Consider an arbitrary vector $\boldsymbol r=[r_1,r_2,r_3]$ fixed in the body frame rotating with the body. This frame was introduced in Appendix A as $\{\bar{\boldsymbol Y}_1, \bar{\boldsymbol Y}_2, \bar{\boldsymbol Y}_3\}$, but we drop the bar symbol here for simplicity. The quaternion representing this point attached to the body is $\rho=(0,\boldsymbol r)$, and
it is constant in time. The corresponding position vector and quaternion in
the inertial frame attached to the orbit are $\hat{\boldsymbol r}$ and
$\hat\rho=(0,\hat{\boldsymbol r})$. By the rules of quaternion algebra, the rotation from $\boldsymbol r$ to $\hat{\boldsymbol r}$ is performed by the Hamilton product
of quaternions
\eb
\hat\rho = q^{-1}\, \rho\,q.
\label{rho.eq}
\ee
Our goal is to derive time derivatives of $q$. Differentiating by $t$,
\eb
\frac{d\hat\rho}{dt} =  \frac{dq^{-1}}{dt}\,\rho\,q + q^{-1}\, \rho\, \frac{dq}{dt},
\ee
and further substituting from Eq. \ref{rho.eq} the inverse rotation
$ \rho = q\,\hat\rho\,q^{-1}$, we obtain
\eb
\frac{d\hat\rho}{dt} =  \frac{dq^{-1}}{dt}\,q\,\hat\rho +  \hat\rho\,q^{-1}\, \frac{dq}{dt}.
\ee
Because $q$ is a unit quaternion, $q^{-1}\,q =1$, and
\eb
\frac{dq^{-1}}{dt}\,q + q^{-1}\,\frac{dq}{dt} = 0,
\ee
hence,
\eb 
\frac{d\hat\rho}{dt} =  -q^{-1}\,\frac{dq}{dt}\,\hat\rho +\hat\rho\, q^{-1}\frac{dq}{dt}
\label{d1.eq}
\ee

Let 
\eb
u = q^{-1}\frac{dq}{dt} .
\label{d2.eq}
\ee
The inverse unit quaternion $q^{-1}$ equals conjugate $q$, $q^{-1}=(q_0,-q_1,-q_2,-q_3)$. The scalar part of $u$ equals zero,
\eb
u_0= \frac{dq_0}{dt}q_0 +  \frac{dq_1}{dt}q_1 + \frac{dq_2}{dt}q_2 + \frac{dq_3}{dt}q_3 = 0,
\ee
because the norm of $q$ is constant. 
So $u$ is a vector quaternion (as well as $\hat\rho$), thus\footnote{A useful
relation underpinning these these transformations is, for the product of two
quaternions $s$ and $p$, $(s_0,\boldsymbol s)(p_0,\boldsymbol p)=(s_0 p_0-
\boldsymbol s\cdot \boldsymbol p)+(s_0\,\boldsymbol p+p_0\,\boldsymbol s+
\boldsymbol s\times \boldsymbol p$).}
\eb 
-u \,\hat\rho + \hat\rho \, u = -2 \,\boldsymbol u \times \hat{\boldsymbol r}
\label{w.eq}
\ee
(using $\times$ as a sign of vector product).

The time derivative of vector $\hat{\boldsymbol r}$, by definition, is 
\eb 
\frac{d\hat{\boldsymbol r}}{dt} = \hat{\boldsymbol \omega} \times \hat{\boldsymbol r}. 
\ee
From Eqs. \ref{d1.eq}, \ref{d2.eq}, and \ref{w.eq},
\eb 
\hat\omega = -2\,u = -2 \,q^{-1}\,\frac{dq}{dt}.
\ee
In our integrations, we operate with the angular velocity vector $\boldsymbol\omega$ in the {\it instantaneous inertial} body frame $\{\boldsymbol Y_1,\boldsymbol Y_2,\boldsymbol Y_3\}$ (cf. Sect. \ref{int.sec}) whose quaternion
$(0,\boldsymbol \omega)$ is $\omega=q\,\hat\omega\,q^{-1}$. This obtains
\eb 
\frac {dq} {dt} = -\frac 1 2 \omega\,q.
\label{dqdt.eq}
\ee
In space engineering and maritime applications, it is customary to employ the inverse
quaternion $q^{-1}$, which represents rotation from the body frame to the inertial (world) frame. For example, it is needed to compute the direction of a transmitting antenna attached to a space craft. The time derivative of this quaternion is readily
obtained from Eq. \ref{dqdt.eq} by conjugation:
\eb
\frac {dq^{-1}} {dt} =  \frac 1 2 q^{-1}\,\omega.
\ee

For a driven rotation integration in the quaternion paradigm, the three
Euler equations \ref{eu.eq} should be 
complemented with four additional ODEs that follow from Eq. \ref{dqdt.eq}\footnote{In advanced languages such as Mathematica and Julia,
standard quaternion multiplication can be directly used}:
\begin{eqnarray}
\dot q_0 & = & -0.5\,(q_1 \dot y_1+  q_2 \dot y_2+ q_3 \dot y_3) \nonumber \\
\dot q_1 & = & +0.5\,(q_0 \dot y_1-  q_3 \dot y_2+ q_2 \dot y_3) \nonumber \\
\dot q_2 & = & +0.5\,(q_3 \dot y_1+  q_0 \dot y_2- q_1 \dot y_3) \nonumber \\
\dot q_3 & = & +0.5\,(-q_2 \dot y_1+  q_1 \dot y_2+ q_0 \dot y_3).
\label{dotq.eq}
\end{eqnarray}

The system of 7 ODEs requires 7 initial conditions to be integrated if its order is reduced to one. This is achieved by explicitly using the vector of angular velocity $\boldsymbol \omega=[\dot y_1,\dot y_2,\dot y_3]$ in the body frame thus eliminating the second time derivatives of the angles $y_i$. The initial conditions then define the initial angular velocity $\boldsymbol \omega(0)=[\dot y_1(0),\dot y_2(0),\dot y_3(0)]$ and the components of the initial attitude quaternion $q=(q_0,q_1,q_2,q_3)$. If the
starting time $t=0$ corresponds to a pericenter time, the initial quaternion defines
how much the body is tilted and inclined with respect to the equilibrium orientation
when the longest axis of the ellipsoid is aligned with the direction to the perturber and the equator is in the orbit plane.
A useful geometric interpretation is that a rotation by angle $\phi$ around a unit
vector $\boldsymbol h$ is represented by a quaternion $(\cos(\phi/2), \sin(\phi/2)\,\boldsymbol h)$.

\end{document}